\documentclass[twocolumn]{jpsj2}
\usepackage{graphicx}
\usepackage{amsmath,amssymb}

\def\runauthor{Youichi {\sc Yanase} and Manfred {\sc Sigrist}}

\title{ 
Superconductivity and Magnetism in Non-centrosymmetric System: 
\\ Application to CePt$_3$Si
} 

\author{Youichi {\sc Yanase}$^{1,2}$\footnote{E-mail:
yanase@itp.phys.ethz.ch} and Manfred {\sc Sigrist}$^{2,3}$}

\inst{1 Department of Physics, University of Tokyo, Tokyo 113-0033, Japan \\
\noindent 2 Theoretische Physik, ETH-Honggerberg, 8093 Zurich, Switzerland \\
\noindent 3 Department of Physics, Kyoto University, Kyoto 606-8502, Japan}

\recdate{Today 2008}

\abst
{ 
 Superconductivity and magnetism in the non-centrosymmetric heavy fermion 
compound CePt$_3$Si and related materials are theoretically investigated. 
 On the basis of the random phase approximation (RPA) analysis 
of the extended Hubbard model, 
we describe the helical spin fluctuation induced by 
the Rashba-type anti-symmetric spin-orbit coupling 
and identify two stable superconducting phases  
with either the dominant $p$-wave ($s$+$P$-wave) symmetry or 
the $d$-wave ($p$+$D$+$f$-wave) symmetry. 
 The effect of the coexistent antiferromagnetic order is 
investigated in both states. 
 The superconducting order parameter, quasiparticle density of state, 
NMR $1/T_{1}T$, specific heat, anisotropy of $H_{\rm c2}$, and possible 
multiple phase transitions are discussed in detail. 
 A comparison with experimental results indicates that 
the $s$+$P$-wave superconducting state is likely realized in CePt$_3$Si. 
}

\kword
{
superconductivity without inversion center, anti-ferromagnetic superconductor, 
helical spin fluctuation 
}

\begin{document}
\sloppy
\maketitle

\newcommand{\eli}{$\acute{{\rm E}}$liashberg }
\renewcommand{\k}{\vec{k}}
\renewcommand{\d}{\vec{d}}
\newcommand{\kk}{\vec{k}'}
\newcommand{\kp}{\vec{k}_{+}}
\newcommand{\kkk}{\vec{k}''}
\newcommand{\q}{\vec{q}}
\newcommand{\Q}{\vec{Q}}
\newcommand{\qp}{\vec{q}_{+}}
\newcommand{\e}{\varepsilon}
\newcommand{\ee}{e}
\newcommand{\s}{{\mit{\it \Sigma}}}
\newcommand{\J}{\mbox{\boldmath$J$}}
\newcommand{\vv}{\mbox{\boldmath$v$}}
\newcommand{\Jh}{J_{{\rm H}}}
\newcommand{\LL}{\mbox{\boldmath$L$}}
\renewcommand{\SS}{\mbox{\boldmath$S$}}
\newcommand{\Tc}{$T_{\rm c}$ }
\newcommand{\Tcf}{$T_{\rm c}$}
\newcommand{\Hc}{$H_{\rm c2}$ }
\newcommand{\Hcf}{$H_{\rm c2}$}
\newcommand{\etal}{{\it et al.}: }
\newcommand{\Pt}{CePt$_3$Si }
\newcommand{\Rh}{CeRhSi$_3$ }
\newcommand{\Ir}{CeIrSi$_3$ }
\newcommand{\Ptf}{CePt$_3$Si}
\newcommand{\Rhf}{CeRhSi$_3$}
\newcommand{\Irf}{CeIrSi$_3$}
\newcommand{\PRL}{Phys. Rev. Lett. } 
\newcommand{\PRB}{Phys. Rev. B } 
\newcommand{\JPSJ}{J. Phys. Soc. Jpn. } 
\newcommand{\Neel}{Ne\'el }

\section{Introduction}

The discovery of superconductivity in materials without an inversion 
center~\cite{rf:bauerDC,rf:bauerreview}  
has initiated intensive research on a new aspect of unconventional
superconductivity. Several new non-centrosymmetric superconductors 
with unique properties have been identified among heavy fermion systems 
such as \Ptf,~\cite{rf:bauerDC,rf:bauerreview} 
UIr,~\cite{rf:akazawa} CeRhSi$_3$,~\cite{rf:kimura,rf:kimurareview} 
CeIrSi$_3$,~\cite{rf:sugitani,rf:onukireview} 
CeCoGe$_3$~\cite{rf:CeCoGe} and others like 
Li$_2$Pd$_{x}$Pt$_{3-x}$B,~\cite{rf:togano} 
Y$_2$C$_3$,~\cite{rf:akimitsu}
Rh$_2$Ga$_9$, Ir$_2$Ga$_9$,~\cite{rf:shibayama}
Mg$_{10}$Ir$_{19}$B$_{16}$,~\cite{rf:mu,rf:klimczuk}
Re$_3$W,~\cite{rf:zuev} boron-doped SiC~\cite{rf:kreiner}
and some organic materials.~\cite{rf:ohmichi}

 One immediate consequence of non-centrosymmetricity is
the necessity for a revised classification scheme of Cooper pairing 
states, as parity is not available as a distinguishing symmetry. 
 Superconducting (SC) states are considered as a mixture of pairing 
states with different parities or, equivalently, the spin configuration is 
composed of both a singlet component and a triplet component. 
 The mixing of spin singlet and spin triplet pairings is induced by the 
anti-symmetric spin-orbit coupling (ASOC).~\cite{rf:edelsteinMixChi} 
 Recent theoretical studies led to the discussion of 
interesting properties of a non-centrosymmetric superconductor, such as the 
magnetelectric effect,~\cite{rf:edelsteinMEE,rf:fujimotoChi,rf:fujimoto,
rf:yip} 
anisotropic spin susceptibility~\cite{rf:edelsteinMixChi,rf:fujimotoChi,
rf:samokhinChi,rf:gorkov,rf:frigeri,rf:frigeriChi,rf:yanaselett,rf:mineevChi,
rf:bulaevskii,rf:yanasehelical} accompanied by the anomalous paramagnetic 
depairing effect,~\cite{rf:yanasehelical} 
anomalous coherence factor in NMR $1/T_{1}T$,~\cite{rf:fujimoto,rf:hayashiT1} 
anisotropic SC gap,~\cite{rf:fujimotoGap,rf:hayashiSD,rf:eremin,
rf:yanaselett,rf:hayashiT1} 
helical SC phase,~\cite{rf:samokhinHelical,rf:kaurHV,rf:agterberg,rf:oka,
rf:mineevHelical,rf:feigelman,rf:yanasehelical} 
Fulde-Ferrel-Larkin-Ovchinnikov (FFLO) state at zero magnetic 
field,~\cite{rf:tanaka} 
various impurity effects,~\cite{rf:mineevImp,rf:frigeriImp,rf:samokhinImp,
rf:ereminImp} 
vortex state,~\cite{rf:nagai,rf:matsunaga} 
and tunneling/Josephson effect.~\cite{rf:yokoyama,rf:sudbo,
rf:hayashiJP,rf:sergienko,rf:iniotakis,rf:varma}

 Non-centrosymmetric heavy fermion superconductors, i.e., 
CePt$_3$Si, UIr, CeRhSi$_3$, CeIrSi$_3$, and CeCoGe$_3$, are of 
particular interests because non-$s$-wave superconductivity is realized owing to
strong electron correlation effects and magnetism has an important effect on 
the superconducting phase. 
 However, the relation between magnetism and superconductivity 
has not been theoretically studied so far, except in studies  
refs.~28, 53, and 54. 
 Here, we extend our previous study~\cite{rf:yanaselett} and investigate 
the pairing state arising from magnetic fluctuation in detail.

 Another aim of this study is to elucidate the effects of 
antiferromagnetic (AFM) order on the SC phase. 
 Interestingly, all presently known non-centrosymmetric heavy fermion 
superconductors
coexist with magnetism.  
 We have shown that some unique properties of \Pt at ambient pressure 
can be induced by the AFM order.~\cite{rf:yanaselett,rf:yanasehelical} 
 In this study, we analyze this issue in more detail.

 Among non-centrosymmetric heavy fermion superconductors, 
CePt$_3$Si has been investigated most extensively because 
its superconductivity occurs at ambient pressure;~\cite{rf:bauerDC} 
others superconduct  only under substantial pressure. 
 Therefore, we focus here on CePt$_3$Si. 
 We believe that 
some of our results are qualitatively valid for other compounds too. 
 In CePt$_3$Si, superconductivity with $T_{\rm c} \sim 0.5$K appears 
in the AFM state with a \Neel temperature 
$T_{\rm N}=2.2$K.~\cite{rf:bauerDC} 
 The AFM order microscopically coexists with 
superconductivity.~\cite{rf:amatoCePtSi,rf:yogiT1}
 Neutron scattering measurements characterize the AFM order 
with an ordering wave vector $\vec{Q}=(0,0,\pi)$ and magnetic moments in the 
{\it ab}-plane of a tetragonal crystal lattice.~\cite{rf:metoki} 
 The AFM order is suppressed by pressure and vanishes at 
a critical pressure $P_{\rm c} \sim 0.6$GPa. 
 Superconductivity is more robust against pressure and therefore 
a purely SC phase is present above the critical pressure 
$P > 0.6$GPa.~\cite{rf:yasuda,rf:tateiwa,rf:takeuchiP}

 The nature of the SC phase has been clarified by several experiments. 
 The low-temperature properties of thermal conductivity,~\cite{rf:izawa} 
superfluid density,~\cite{rf:bonalde} specific heat,~\cite{rf:takeuchiC} 
and NMR $1/T_{1}T$~\cite{rf:mukudaT1} indicate line nodes in the gap. 
 The upper critical field $H_{\rm c2} \sim 3 - 4$T exceeds  the standard 
paramagnetic limit~\cite{rf:bauerDC}, which seems to be consistent with the 
Knight shift data displaying no decrease in spin susceptibility 
below $T_{\rm c}$ for any field direction.~\cite{rf:yogiK,rf:higemoto} 
 The combination of these features is incompatible with the usual 
pairing states such as the $s$-wave, $p$-wave, or $d$-wave state, 
and calls for an extension of the standard working scheme.

 In ref.~29, we have investigated the magnetic properties 
of non-centrosymmetric superconductors.  
 Then, it was shown that the predominantly $p$-wave state 
admixed with the $s$-wave order parameter ($s$+$P$-wave state) 
is consistent with the paramagnetic properties of \Ptf. 
 We here examine the symmetry of superconductivity in \Pt from 
the microscopic point of view and show that the $s$+$P$-wave state 
or $p$+$D$+$f$-wave state can be stabilized by spin fluctuation 
with helical anisotropy. 
 We also calculate the quasiparticle excitations, specific heat, and  
NMR $1/T_{1}T$, and show that the line node behavior in \Pt at ambient 
pressure is consistent with the $s$+$P$-wave state. 
 We investigate the pressure dependence of these quantities, possible 
SC multiple phase transitions, and the anisotropy of \Hcf. 
Some future experimental tests are proposed.

 The paper is organized as follows. 
 In \S2, we formulate the RPA theory in the Hubbard model with ASOC 
and AFM order.  
 The nature of spin fluctuation and superconductivity is investigated 
in \S3 and \S4, respectively. 
 The symmetry of superconductivity, SC gap structure, specific heat and NMR 
$1/T_{1}T$, multiple SC phase transitions, and anisotropy of \Hc 
are discussed in \S4.1, \S4.2, \S4.3, \S4.4 and \S4.5, respectively. 
 Some future experiments are proposed in \S4. 
 The summary and discussions are given in \S5. 
 A derivation of ASOC in the periodical Anderson model and Hubbard model 
is given in Appendix.

\section{Formulation}

\subsection{Hubbard model with ASOC and AFM order}

 For the following study of superconductivity in CePt$_3$Si, 
we introduce the single-orbital Hubbard model including the AFM order 
and ASOC 
\begin{eqnarray}
\label{eq:Hubbard-model}
&& \hspace*{-7mm}  H = \sum_{k,s} \e(\k) c_{\k,s}^{\dag}c_{\k,s} 
   + \alpha  \sum_{k} \vec{g}(\k) \cdot \vec{S}(\k)  
   - \sum_{k} \vec{h}_{\rm Q}  \cdot \vec{S}_{\rm Q}(\k) 
\nonumber \\
&& \hspace*{-0mm}   
   + U \sum_{i} n_{i,\uparrow} n_{i,\downarrow}, 
%
\end{eqnarray}
where $\vec{S}(\k) = \sum_{s,s'} \vec{\sigma}_{ss'} c_{\k,s}^{\dag}c_{\k,s'}$ 
and $\vec{S}_{\rm Q}(\k) = 
\sum_{s,s'} \vec{\sigma}_{ss'} c_{\k+\Q,s}^{\dag}c_{\k,s'}$ with 
$\vec{\sigma}_{ss'}$ being the vector representation of the Pauli matrix. 
 $n_{i,s}$ is the electron number at the site $i$ with the spin $s$. 
 We do not touch the heavy Fermion aspect, 
i.e., the hybridization of conduction electrons with Ce 4$f$-electrons 
forming strongly renormalized quasiparticles. However, we consider the 
Hubbard model as a valid effective model for describing low-energy 
quasiparticles in the Fermi liquid state.~\cite{rf:yanaseReview}

 We consider a simple tetragonal lattice and assume the 
dispersion relation as 
\begin{eqnarray}
\label{eq:dispersion}
&& \hspace*{-8mm}  \e(\k)  =   2 t_1 (\cos k_{\rm x} +\cos k_{\rm y}) 
         + 4 t_2 \cos k_{\rm x} \cos k_{\rm y} 
\nonumber \\ && \hspace*{-8mm} 
         + 2 t_3 (\cos 2 k_{\rm x} +\cos 2 k_{\rm y})
+ [ 2 t_4 + 4 t_5 (\cos k_{\rm x} +\cos k_{\rm y}) 
\nonumber \\ && \hspace*{-8mm} 
         + 4 t_6 (\cos 2 k_{\rm x} +\cos 2 k_{\rm y}) ] \cos k_{\rm z}
         + 2 t_7 \cos 2 k_{\rm z} 
         - \mu, 
\end{eqnarray}
where the chemical potential $\mu$ is included. 
 We determine the chemical potential $\mu$ so that the electron density per 
site is $n$. 
 By choosing the parameters as 
$(t_1,t_2,t_3,t_4,t_5,t_6,t_7,n) = (1,-0.15,-0.5,-0.3,-0.1,-0.09,-0.2,1.75)$, 
the dispersion relation eq.~(2) reproduces 
the $\beta$-band of CePt$_3$Si, which has been reported 
by band structure calculation without the AFM 
order.~\cite{rf:samokhinband,rf:anisimov,rf:hashimotoDHV} 
The Fermi surface of this tight-binding model is depicted in 
Fig.~1 of ref.~28. 
We assume that the superconductivity in \Pt is mainly 
induced by the $\beta$-band because the $\beta$-band has a substantial 
Ce 4$f$-electron character~\cite{rf:anisimov} and the largest density of 
states (DOS), namely 70\% of the total DOS.~\cite{rf:samokhinband}

 The second term in eq.~(1) describes the ASOC that arises from the lack of 
inversion symmetry and is characterized by the vector $\vec{g}(\k)$. 
 Time reversal symmetry is preserved, if the $g$-vector 
is odd in $\k$, i.e., $\vec{g}(-\k)=-\vec{g}(\k)$. 
 In the case of \Pt as well as of \Rh and \Irf, 
the $g$-vector has the Rashba type structure.~\cite{rf:rashba} 
 The microscopic derivation of the ASOC in the $f$-electron systems 
is given in Appendix. The ASOC in the periodic Anderson model 
as well as that in the Hubbard model originate from the combination of 
the atomic $L$-$S$ coupling in the $f$-orbital and the hybridization with 
conduction electrons. 
 Although the detailed momentum dependence of the $g$-vector 
is complicated (see eq.~(A.25)) and is difficult to obtain by band structure 
calculations, at least from a symmetry point of view,  
$\vec{g}(\k)= (- v_{\rm y}(\k) , v_{\rm x}(\k), 0) /\bar{v}$ delivers 
a reasonable approximation, 
where $v_{\rm x,y}(\k) = \partial \e(\k)/\partial k_{\rm x,y}$ 
is the quasiparticle velocity. 
 We normalize  $\vec{g}(\k) $ by the average velocity $\bar{v}$ 
[$\bar{v}^{2}=\frac{1}{N}\sum_{k}v_{\rm x}(\k)^{2}+v_{\rm y}(\k)^{2}$] 
so that the coupling constant $\alpha$ has the dimension of energy. 
 This form reproduces the symmetry and periodicity of 
the Rashba-type $g$-vector within the Brillouin zone. 
 We choose the coupling constant $\alpha=0.3$ in the main part of this paper 
so that the band splitting due to ASOC is consistent with the band structure 
calculations.~\cite{rf:samokhinband}

 The AFM order enters in our model through the staggered field 
$\vec{h}_{\rm Q}$ without discussing its microscopic origin. 
 The phase diagram under pressure implies that 
the AFM order mainly arises from 
localized Ce 4$f$-electrons that have a 
character different from that of SC quasiparticles. 
 The \Tc of superconductivity is slightly affected by the AFM order which 
vanishes at $P \sim 0.6$GPa,~\cite{rf:yasuda,rf:tateiwa,rf:takeuchiP} 
in contrast to the other Ce-based superconductors.~\cite{rf:kitaoka} 
 The experimentally determined AFM order corresponds to $ \vec{h}_{\rm Q} 
=h_{\rm Q} \hat{x}$ pointing in the [100] direction with a wave vector 
$ \Q = (0,0,\pi) $.~\cite{rf:metoki}  
 For the magnitude, we choose $|h_{\rm Q}| \ll W $ where $W$ is the bandwidth 
since the observed AFM moment $\sim 0.16 \mu_{\rm B}$ is considerably less 
than the full moment of the $5/2$ manifold in 
the Ce ion.~\cite{rf:metoki}

 The undressed Green functions for $U=0$ are represented by the matrix form  
$
\hat{G}(\k,{\rm i}\omega_{n}) = 
({\rm i}\omega_{n} \hat{1} - \hat{H}(\k))^{-1},
$
where 
\begin{eqnarray}
\label{eq:Green-function}
\hat{G}(\k,{\rm i}\omega_{n}) = 
\left(
\begin{array}{cc}
\hat{G}^{1}(\k,{\rm i}\omega_{n}) & 
\hat{G}^{2}(\k,{\rm i}\omega_{n})\\
\hat{G}^{2}(\kp,{\rm i}\omega_{n}) & 
\hat{G}^{1}(\kp,{\rm i}\omega_{n})\\
\end{array}
\right),  
\end{eqnarray}
and
\begin{eqnarray}
\label{eq:H-matrix}
\hat{H}(\k)
= 
\left(
\begin{array}{cc}
\hat{e}(\k) &
-h_{\rm Q} \hat{\sigma}^{({\rm x})} \\
-h_{\rm Q} \hat{\sigma}^{({\rm x})} & 
\hat{e}(\kp) \\
\end{array}
\right), 
\end{eqnarray}
%
with $\hat{e}(\k)=\e(\k)\hat{\sigma}^{(0)} 
+ \alpha \vec{g}(\k) \vec{\sigma}$ and $\kp = \k+\Q$.  
 The normal and anomalous Green functions $\hat{G}^{i}(\k,{\rm i}\omega_{n})$ 
are the 2 $\times$ 2 matrix in spin space, where $\omega_{n}=(2 n + 1) \pi T$ 
and $T$ is the temperature.

\subsection{$\acute{E}$liashberg equation}

 We turn to the SC instability that we assume to arise through
electron-electron interaction incorporated in the effective on-site 
repulsion $U$. 
 The linearized \eli equation is obtained by the standard procedure: 
\begin{eqnarray}
\label{eq:Eliashberg}
&& \hspace*{-11mm} \lambda \Delta_{p,s_1,s_2} (\k) = 
\nonumber \\ && \hspace*{0mm} 
- \sum_{\rm k',q,s_3,s_4} V_{p,q,s_1,s_2,s_3,s_4}(\k,\kk) \psi_{q,s_3,s_4}(\kk),
\\
&& \hspace*{-11mm} 
\psi_{p,s_1,s_2}(\k) = \sum_{i,j,s_3,s_4} \phi_{p,i,j,s_1,s_3,s_2,s_4}(\k) 
\Delta_{q,s_3,s_4}(\kkk),
\end{eqnarray}
where $q=p$ ($q=3-p$) for $i=j$ ($i \ne j$), $\kkk=\k+(i-1)\Q$ and 
\begin{eqnarray}
&& \hspace*{-10mm}  
\phi_{p,i,j,s_1,s_2,s_3,s_4}(\k) =  
T \sum_{n} G^{i}_{s_1,s_2}(\k,{\rm i}\omega_{n})  
\nonumber \\ && \hspace*{3mm} 
\times 
           G^{j}_{s_3,s_4}(-\k + (p-1)\Q,-{\rm i}\omega_{n}) 
\hspace*{3mm} (p=1,2). 
\end{eqnarray}
 Here, we adopt the so-called weak coupling theory of superconductivity 
and ignore self-energy corrections and the frequency dependence 
of effective interaction.~\cite{rf:yanaseReview,rf:miyake}
 This simplification strongly affects the resulting transition temperature 
but hardly affects the symmetry of pairing.~\cite{rf:yanaseReview} 
We denote the order parameter for the superconductivity as  
$\Delta_{p,s_1,s_2} (\k)$ ($p=1,2$, $s_1$ and $s_2$ are the spin indices), where $\Delta_{1,s_1,s_2} (\k)$ and 
$\Delta_{2,s_1,s_2} (\k)$ describe the Cooper pairing with the total momenta 
$(0,0,0)$ and $(0,0,\pi)$, respectively. 
 The former is the order parameter for ordinary Cooper pairs, 
while the latter is that for $\pi$-singlet and $\pi$-triplet pairs. 
 These $\pi$-pairs are admixed with 
usual Cooper pairs in the presence of the AFM order.~\cite{rf:pi-triplet}

 The effective interaction 
$V_{p,q,s_1,s_2,s_3,s_4}(\k,\kk)$ originates 
from spin fluctuations that we describe within the RPA~\cite{rf:miyake} 
according to the diagrammatic expression shown in Fig.~1. 
 In the RPA, the effective interaction is described by the generalized 
susceptibility whose matrix form is expressed as  
\begin{eqnarray}
\label{eq:kai}
&& \hspace*{-10mm}
\left(
\begin{array}{c}
\hat{\chi}_{1}(\q) \\
\hat{\chi}_{2}(\qp) \\
\end{array}
\right) 
=
\nonumber \\ && \hspace*{-14mm} 
\left(
\begin{array}{cc}
\hat{1} - \hat{\chi}_{1}^{(0)}(\q) \hat{U} &  
- \hat{\chi}_{2}^{(0)}(\q) \hat{U} \\
- \hat{\chi}_{2}^{(0)}(\qp) \hat{U} & 
\hat{1} - \hat{\chi}_{1}^{(0)}(\qp) \hat{U} \\
\end{array}
\right)^{-1}
\left(
\begin{array}{c}
\hat{\chi}_{1}^{(0)}(\q) \\
\hat{\chi}_{2}^{(0)}(\qp) \\
\end{array}
\right), 
\end{eqnarray}
where $\qp=\q+\Q$. 
Hereafter, we denote the element of the $4 \times 4$ matrix, such as 
$\hat{A}=\hat{\chi}_{i}^{(0)}(\q)$ and $\hat{\chi}_{i}(\q)$ 
using the spin indices as  
\begin{eqnarray}
\label{eq:kaimatrix}
&& \hspace*{-5mm} 
\hat{A}
=
\left(
\begin{array}{cccc}
A_{\uparrow\uparrow\uparrow\uparrow} &
A_{\uparrow\uparrow\uparrow\downarrow} &
A_{\uparrow\uparrow\downarrow\uparrow} &
A_{\uparrow\uparrow\downarrow\downarrow} \\
A_{\uparrow\downarrow\uparrow\uparrow} &
A_{\uparrow\downarrow\uparrow\downarrow} &
A_{\uparrow\downarrow\downarrow\uparrow} &
A_{\uparrow\downarrow\downarrow\downarrow} \\
A_{\downarrow\uparrow\uparrow\uparrow} &
A_{\downarrow\uparrow\uparrow\downarrow} &
A_{\downarrow\uparrow\downarrow\uparrow} &
A_{\downarrow\uparrow\downarrow\downarrow} \\
A_{\downarrow\downarrow\uparrow\uparrow} &
A_{\downarrow\downarrow\uparrow\downarrow} &
A_{\downarrow\downarrow\downarrow\uparrow} &
A_{\downarrow\downarrow\downarrow\downarrow} \\
\end{array}
\right). 
\end{eqnarray}
%
%
%
 The matrix element of the bare susceptibility 
$\hat{\chi}_{i}^{(0)}(\q)$ is expressed as 
\begin{eqnarray}
\label{eq:kaizero}
&& \hspace{-10mm}
\chi_{1,s1,s2,s3,s4}^{(0)}(\q) 
=
\nonumber \\ && \hspace*{0mm} 
-T\sum_{\k,\omega_{\rm n}}
[
G^{1}_{s4,s1}(\k+\q,{\rm i}\omega_{n}) 
G^{1}_{s2,s3}(\k,{\rm i}\omega_{n}) 
\nonumber \\ && \hspace*{0mm} 
+
G^{2}_{s4,s1}(\k+\q,{\rm i}\omega_{n}) 
G^{2}_{s2,s3}(\k+\Q,{\rm i}\omega_{n})
], 
\\
&& \hspace{-10mm}
\chi_{2,s1,s2,s3,s4}^{(0)}(\q) 
=
\nonumber \\ && \hspace*{0mm} 
-T\sum_{\k,\omega_{\rm n}}
[
G^{1}_{s4,s1}(\k+\q,{\rm i}\omega_{n}) 
G^{2}_{s2,s3}(\k,{\rm i}\omega_{n}) 
\nonumber \\ && \hspace*{0mm} 
+
G^{2}_{s4,s1}(\k+\q,{\rm i}\omega_{n}) 
G^{1}_{s2,s3}(\k+\Q,{\rm i}\omega_{n})
], 
\end{eqnarray}
and the matrix $\hat{U}$ is obtained as 
\begin{eqnarray}
\label{eq:interaction}
\hat{U}
=
\left(
\begin{array}{cccc}
0 & 0 & 0 & -U \\
0 & 0 & U & 0 \\
0 & U & 0 & 0 \\
-U & 0 & 0 & 0 \\
\end{array}
\right). 
\end{eqnarray}
 According to Fig.~1, the effective interaction is obtained as  
\begin{eqnarray}
\label{eq:effective}
&& \hspace*{-16mm} 
V_{1,1,s1,s2,s3,s4}(\k,\kk) = V_{2,2,s1,s2,s3,s4}(\k,\kk)
\nonumber \\ && \hspace*{-5mm} 
= - [\hat{U'} \hat{\chi}_{1}(\kk-\k) \hat{U'}]_{s3,s1,s4,s2}
+ \hat{U}_{s1,s2,s3,s4}, 
\\
&& \hspace*{-16mm} 
V_{1,2,s1,s2,s3,s4}(\k,\kk) = V_{2,1,s1,s2,s3,s4}(\k,\kk)
\nonumber \\ && \hspace*{-5mm} 
= - [\hat{U'} \hat{\chi}_{2}(\kk-\k) \hat{U'}]_{s3,s1,s4,s2}, 
\end{eqnarray}
where 
\begin{eqnarray}
\label{eq:interactions}
\hat{U'}
=
\left(
\begin{array}{cccc}
0 & 0 & 0 & -U \\
0 & U & 0 & 0 \\
0 & 0 & U & 0 \\
-U & 0 & 0 & 0 \\
\end{array}
\right). 
\end{eqnarray}
%

\begin{figure}[htbp]
  \begin{center}
\includegraphics[width=8cm]{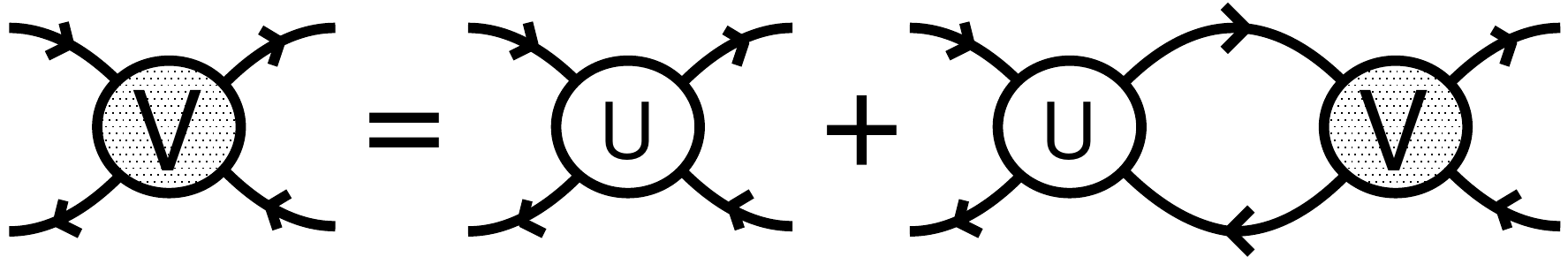}
\caption{
Diagrammatic representation of the pairing interaction 
$V_{p,q,s1,s2,s3,s4}(\k,\kk)$. 
The white circle represents the on-site interaction $U$. 
}
    \label{fig:diagrams}
  \end{center}
\end{figure}

 The linearized \eli equation (eqs.~(5)-(7)) allows us to determine 
the form of the leading pairing instability, 
which is attained for the temperature at which the largest eigenvalue 
$ \lambda $ reaches unity. 
 Numerical accuracy requires, however, a different but equivalent approach. 
 We perform the calculation at a given temperature, in our case $T=0.02$,  
which is much lower than the Fermi temperature, 
and determine the most stable pairing state as the eigenfunction of 
the largest eigenvalue.~\cite{rf:yanaseReview} 
 The typical eigenvalue at $T=0.02$ and $U=4$ lies at around 
$\lambda = 0.3 \sim 0.6$. 
 This means that the \Tc for $U=4$ is lower than $T=0.02$. 
 However, the absolute value of \Tc is not important for our purpose,  
which is focused on the roles of the spin fluctuation, ASOC, and AFM order. 
 We believe that the qualitative roles of these aspects can be captured 
in this simple calculation. 
 On the other hand, the absolute value of \Tc is significantly affected 
by the mode coupling effect, vertex corrections, strong coupling effect, 
and multi-orbital effect, which are neglected in our calculation. 
 We leave more sophisticated calculation based on the multi-orbital model 
and beyond the RPA for future discussion.

 The \Tc of superconductivity reaches $T_{\rm c}=0.02$ 
if we assume a larger $U$. However, we show the results 
for $U=4$, unless stated otherwise. 
 This is mainly because the results for a large $U$ are likely spurious
because of the limitation of RPA. 
 Since the mode coupling effect is neglected in the RPA, 
the critical fluctuation is not taken into account and therefore the 
magnetic instability is seriously overestimated. 
 When we assume a large $U$ so that we obtain a high \Tcf, the 
system approaches  the magnetic instability, which is beyond the 
applicability of RPA. 
 Our results for the superconductivity are only weakly dependent on $U$ 
except for the relative stability between the $s$+$P$-wave and 
$p$+$D$+$f$-wave states, and therefore we qualitatively obtain 
the same results for a larger $U$. 
 However, we avoid parameters close to the 
magnetic instability.

 Note that we here consider the spin fluctuation arising from  
quasiparticles that are mainly superconducting and may be different 
from the main source of the AFM moment. 
 Although the quantum critical point of the AFM order exists at 
$P \sim 0.6$GPa, the critical fluctuation of the AFM moment slightly 
affects the \Tc of superconductivity, as indicated by the phase diagram 
in the $P$-$T$ plane.~\cite{rf:yasuda,rf:tateiwa,rf:takeuchiP} 
 Therefore, it is expected that the critical fluctuations of the AFM moments are 
only weakly coupled to quasiparticles and the superconductivity is 
mainly induced by the residual interaction between quasiparticles.

\section{Spin Fluctuation}

 First, we investigate the spin fluctuation 
in the non-centrosymmetric system. 
 To clarify the role of ASOC, we consider the paramagnetic state
where $h_{\rm Q}=0$. 
 Then, the static spin susceptibility is obtained as  
\begin{eqnarray}
\label{eq:kaiobs}
&& \hspace*{-10mm}
\chi^{\mu\nu}(\q)  =  \int_{0}^{\beta} {\rm d}\tau 
                       <T_{\tau} S^{\mu}(\q,\tau) S^{\nu}(-\q)>
\\ && \hspace*{-10mm}
= \sum_{s1,s2,s3,s4} 
\sigma^{\mu}_{s1,s2} \chi_{1,s1,s2,s3,s4}(\q) \sigma^{\nu}_{s3,s4},  
\end{eqnarray}
where $S^{\mu}(\q,\tau)=e^{H \tau} S^{\mu}(\q) e^{-H \tau}$ and 
$S^{\mu}(\q) = 
\sum_{\k, s, s'} \sigma^{\mu}_{ss'} c^{\dag}_{\k+\q s} c_{\k s'}$. 
 We define $\chi_{\rm max}(\q)$ as the maximum eigenvalue of  the
$3 \times 3$ matrix $\chi^{\mu\nu}(\q)$ for $\mu$ and $\nu$.

 In the absence of ASOC, the spin susceptibility is isotropic, 
namely $\chi^{\mu\nu}(\q)=0$ for $\mu \ne \nu$ and 
$\chi^{xx}(\q)=\chi^{yy}(\q)=\chi^{zz}(\q)=\chi_{\rm max}(\q)$. 
 The spin susceptibility has a peak at $\q = (0,0,\pi)$ 
because of the band structure of the $\beta$-band, as shown in 
Fig.~2(a) (dashed line).  
 Thus, the $\beta$-band favors the ferromagnetic spin correlation 
in the {\it ab}-plane and the AFM correlation between the plane. 
 This is the spin structure realized in the AFM state of \Ptf.

\begin{figure}[htbp]
  \begin{center}
\includegraphics[width=7cm]{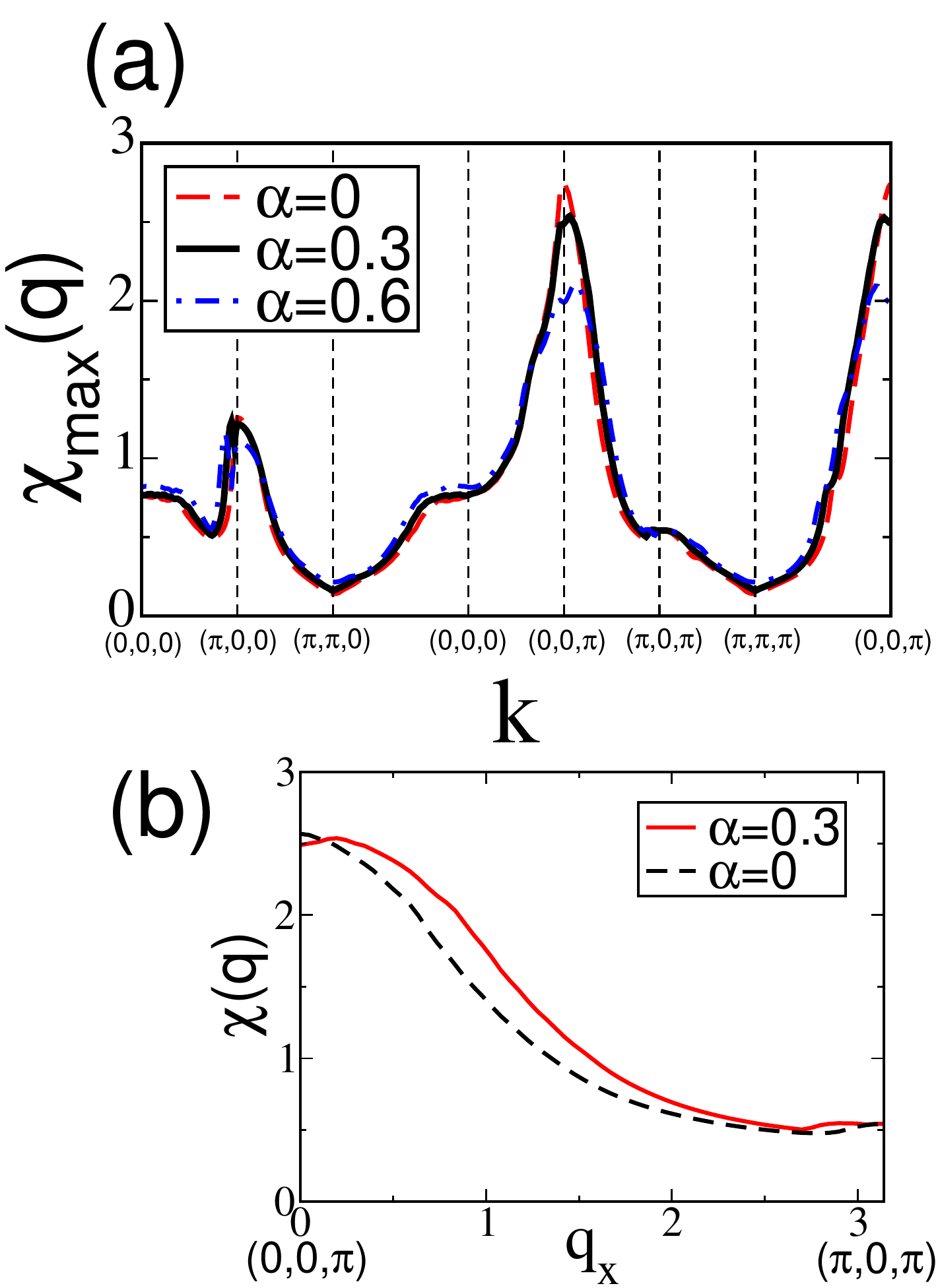}
\caption{(color online)
(a) Momentum dependence of spin susceptibility $\chi_{\rm max}(\q)$ 
in the paramagnetic state ($h_{\rm Q}=0$). 
We assume $U=4$ and $T=0.02$ and show the results for 
$\alpha=0$ (dashed line), $\alpha=0.3$ (solid line) 
and $\alpha=0.6$ (dash-dotted line). 
(b) Spin susceptibility $\chi_{\rm max}(\q)$ along 
$(0,0,\pi)$-$(\pi,0,\pi)$ direction. 
 We assume $U=3.95$ for $\alpha=0$ in (b)
so that the maximum spin susceptibilities are similar 
between $\alpha=0$ and $\alpha=0.3$. 
 Numerical calculation is carried out by dividing the first Brillouin 
zone into a $128 \times 128 \times 48$ lattice. 
}
    \label{fig:spinfluctuation}
  \end{center}
\end{figure}

 The anisotropy of spin susceptibility is induced by the ASOC. 
 Our numerical calculation accurately takes into account the ASOC, 
but we explain here the role of ASOC within the first order of $\alpha$ 
to provide a simple and qualitative understanding of the helical anisotropy. 

 The lowest-order term in $\alpha$ appears in the off-diagonal 
component of the spin susceptibility tensor $\chi^{\mu\nu}(\q)$ 
which is nonzero unless 
$\q_{\parallel} = (n_{\rm x},n_{\rm y}) \pi$ 
with $n_{\rm x}$ and $n_{\rm y}$ integers. 
 The off-diagonal component can be viewed as a result of 
the Dzyaloshinski-Moriya-type interaction $H_{\rm DM}= 
\sum_{\q} {\rm i} D(\q) \cdot S(\q) \times S(-\q)$.~\cite{rf:frigeriImp} 
 Since the Rashba-type ASOC leads to 
$D(\q) \propto \alpha \hat{z} \times \q = \alpha (-q_{\rm y}, q_{\rm x}, 0)$ 
in the vicinity of $\q_{\parallel} = (0,0)$, 
the off-diagonal components of the spin susceptibility tensor 
are described as 
$\chi^{\rm yz}(\q) = -\chi^{\rm zy}(\q)
= {\rm i} \alpha B(q_{\rm z}) q_{\rm y} + O(\alpha^{3})$, 
$\chi^{\rm xz}(\q) = -\chi^{\rm zx}(\q)
= {\rm i} \alpha B(q_{\rm z}) q_{\rm x} + O(\alpha^{3})$, and  
$\chi^{\rm xy}(\q)= \chi^{\rm yx}(\q) = O(\alpha^{2})$. 
 Because the momentum dependence of the diagonal component 
is quadratic as shown by $\chi^{\mu\mu}(\q_{\parallel},q_{\rm z}) 
\sim \chi(0,q_{\rm z}) - A(q_{\rm z}) |\q_{\parallel}|^{2} 
+ O(|\q_{\parallel}|^{4})$, 
the maximum eigenvalue of the spin susceptibility tensor is obtained as 
$\chi_{\rm max}(\q_{\parallel},q_{\rm z}) = 
\chi(0,q_{\rm z}) + \alpha B(q_{\rm z}) |\q_{\parallel}| 
- A(q_{\rm z}) |\q_{\parallel}|^{2} + O(\alpha^{2},|\q_{\parallel}|^{4})$ 
around $\q_{\parallel} = (0,0)$. 
 Thus, $\chi_{\rm max}(\q_{\parallel},q_{\rm z})$ for each $q_{\rm z}$ 
has a local minimum at $\q_{\parallel}=0$ and a local maximum 
at $\q_{\parallel} \ne 0$. 
 For $\alpha=0.3$ the numerical calculation shows four peaks of  
$\chi_{\rm max}(\q_{\parallel},q_{\rm z})$ 
at $\q \sim (0,\pm 0.2,\pi)$ and $\q \sim (\pm 0.2,0,\pi)$ 
in contrast to the single peak at $\q=(0,0,\pi)$ for $\alpha=0$ 
(see Fig.~2(b)).

 Since the off-diagonal components, such as  
$\chi^{\rm xz}(\q)=(\chi^{\rm zx}(\q))^{*}$ 
and 
$\chi^{\rm yz}(\q)=(\chi^{\rm zy}(\q))^{*}$, 
are purely imaginary, the maximum eigenvalue of 
the spin susceptibility tensor has the eigenvector 
$\vec{S}(\q)= \frac{1}{\sqrt2}
(\tilde{q}_{\rm x},\tilde{q}_{\rm y},\pm {\rm i})$ 
with $\tilde{q}_{\rm x,y}=q_{\rm x,y}/|\q_{\parallel}|$.  
 Thus, $\chi_{\rm max}(\q_{\parallel},q_{\rm z})$ describes the 
susceptibility of the helical magnetic order.

 We here discuss the effect of higher-order terms of $\alpha$ by which 
the helical spin structure is distorted. 
 In our result for $\alpha=0.3$, the spin susceptibility tensor at 
$\q =(0.196,0,\pi)$ has a maximum eigenvalue for 
$\vec{S}(\q) \sim (0.81,0,-0.59{\rm i})$. 
 The deviation from $\vec{S}(\q)= \frac{1}{\sqrt2}(1,0,\pm {\rm i})$ 
in the lowest-order theory of $\alpha$ mainly arises from the second-order 
term of $\alpha$ in the diagonal component of $\chi^{\mu\nu}(\q)$.

 Helical magnetism is suppressed by symmetric spin-orbit 
coupling, namely, the atomic $L$-$S$ coupling, which is not taken into 
account in this paper. 
 This is the reason why the helical magnetic order is not actually realized 
in non-centrosymmetric heavy fermion compounds, 
but is observed in non-centrosymmetric compounds with a small 
$L$-$S$ coupling.~\cite{rf:MnSi} 
 However, qualitatively the same effects of the ASOC, such as the 
helical anisotropy of spin susceptibility for $\q_{\parallel} \ne 0$, 
are expected in the presence of $L$-$S$ coupling.

\section{Superconductivity}

\subsection{Pairing symmetry}

 We examine here the superconductivity. 
 First, we discuss the symmetry of the SC state. 
 It is convenient in the following discussions to 
describe the order parameter in a standard 
manner as~\cite{rf:sigrist,rf:leggett} 
\begin{eqnarray}
\label{eq:d-vector}
&& \hspace*{-7mm}
\Delta_{1,s,s'}(\k)
= 
\left(
\begin{array}{cc}
-d_{{\rm x}}(\k)+{\rm i}d_{{\rm y}}(\k) & \Phi(\k) + d_{{\rm z}}(\k) \\
-\Phi(\k) + d_{{\rm z}}(\k) & d_{{\rm x}}(\k)+{\rm i}d_{{\rm y}}(\k) \\
\end{array}
\right), 
\nonumber \\ &&\hspace*{-10mm}
\end{eqnarray}
where we use the even parity scalar function $ \Phi (\k) $ 
and the odd parity $d$-vector $ \vec{d}(\k) $. 
 In the presence of the  AFM order, the order parameter for 
the $\pi$-triplet and $\pi$-singlet pairings 
$\Delta_{2,s,s'}(\k)$ appears owing to the folding of the Brillouin zone.
 However, the basic properties and symmetries are hardly 
affected by $\pi$-parings when $h_{\rm Q} \ll W$.

 We identify two stable solutions of the \eli equation. 
 One pairing state has a predominant $p$-wave symmetry whose order 
parameter has the leading odd parity component 
$\vec{d}(\k) \sim (-\sin k_{\rm y},\beta \sin k_{\rm x}, 0)$. 
  The parameter $\beta$ is unity in the absence of the AFM order. 
 The admixed even parity part is approximated as 
$\Phi(\k) \sim \delta + \cos k_{\rm x}+\cos k_{\rm y}$ 
with $\delta \sim 0.2$. 
 Thus, the spin singlet component has the $s$-wave symmetry, 
as discussed in ref.~24, but its sign changes in the radial 
direction in order to avoid the local repulsive interaction $U$. 
 We denote this pairing state as the $s$+$P$-wave state.

 The other stable solution is the predominantly $d$-wave state 
that can be viewed as an interlayer Cooper pairing state: 
$\Phi(\k) \sim  
\{ \sin k_{\rm x} \sin k_{\rm z} , \sin k_{\rm y} \sin k_{\rm z} \} $ 
(two-fold degenerate) admixed with an odd-parity component  
$\vec{d}(\k) \sim  \Phi(\k) (-\sin k_{\rm y} , \sin k_{\rm x},0 ) $. 
 In the paramagnetic phase, the most stable combination of 
the two degenerate states is chiral: 
$ \Phi_{\pm} (\k) \sim (\sin k_{\rm x} \pm i \sin k_{\rm y}) \sin k_{\rm z} $ 
which gains the maximal condensation energy 
in the weak-coupling approach. 
 In the AFM state, however, the two states of $\Phi(\k)$ are no 
longer degenerate. 
 According the the RPA theory that we adopt in this paper, the 
$d_{\rm xz}$-wave state ($d_{\rm yz}$-wave state) is favored by 
the AFM order along the $\hat{x}$-axis ($\hat{y}$-axis). 
 Since the spin triplet order parameter has both the $p$-wave and 
$f$-wave components, we denote this state as the $p$+$D$+$f$-wave state.

 In the RPA theory, superconductivity is assumed to be induced by the 
spin fluctuation. As discussed in \S3, the spin fluctuation arising from the 
$\beta$-band has four peaks around $\q \sim (0,\pm 0.2,\pi)$ and 
$\q \sim (\pm 0.2,0,\pi)$, which indicates the 
nearly ferromagnetic (helical) spin correlation in the {\it ab}-plane and 
the AFM coupling between the planes. 
 For a small $U$, the spin fluctuation has a two-dimensional nature because of 
the dispersion relation eq.~(2). 
 Then, the interplane AFM coupling is negligible and 
the intraplane nearly ferromagnetic correlation 
induces  $s$+$P$-wave superconductivity. 
 On the other hand, the AFM 
coupling between the planes leads to a three-dimensional spin fluctuation 
for a large $U$ and favors the $p$+$D$+$f$-wave state. 
 Figure~3 shows the phase diagram against $U$ and the AFM staggered field 
$h_{\rm Q}$. 
 We identify the pairing state with the largest eigenvalue in the \eli equation
at $T=0.02$, as mentioned in \S2. 
 We see that the two pairing states are nearly degenerate at around 
$U=3 \sim 3.5$ independently of the AFM staggered field. 

 We find the other pairing state having the predominantly extended 
$s$-wave symmetry with 
$\Phi(\k) \sim \cos k_{\rm x} + \cos k_{\rm y} - 2 \cos k_{\rm z}$ 
as a self-consistent solution of the \eli equation. 
 However, we have found no parameter set where this pairing state 
is stable.

\begin{figure}[htbp]
  \begin{center}
\includegraphics[width=6cm]{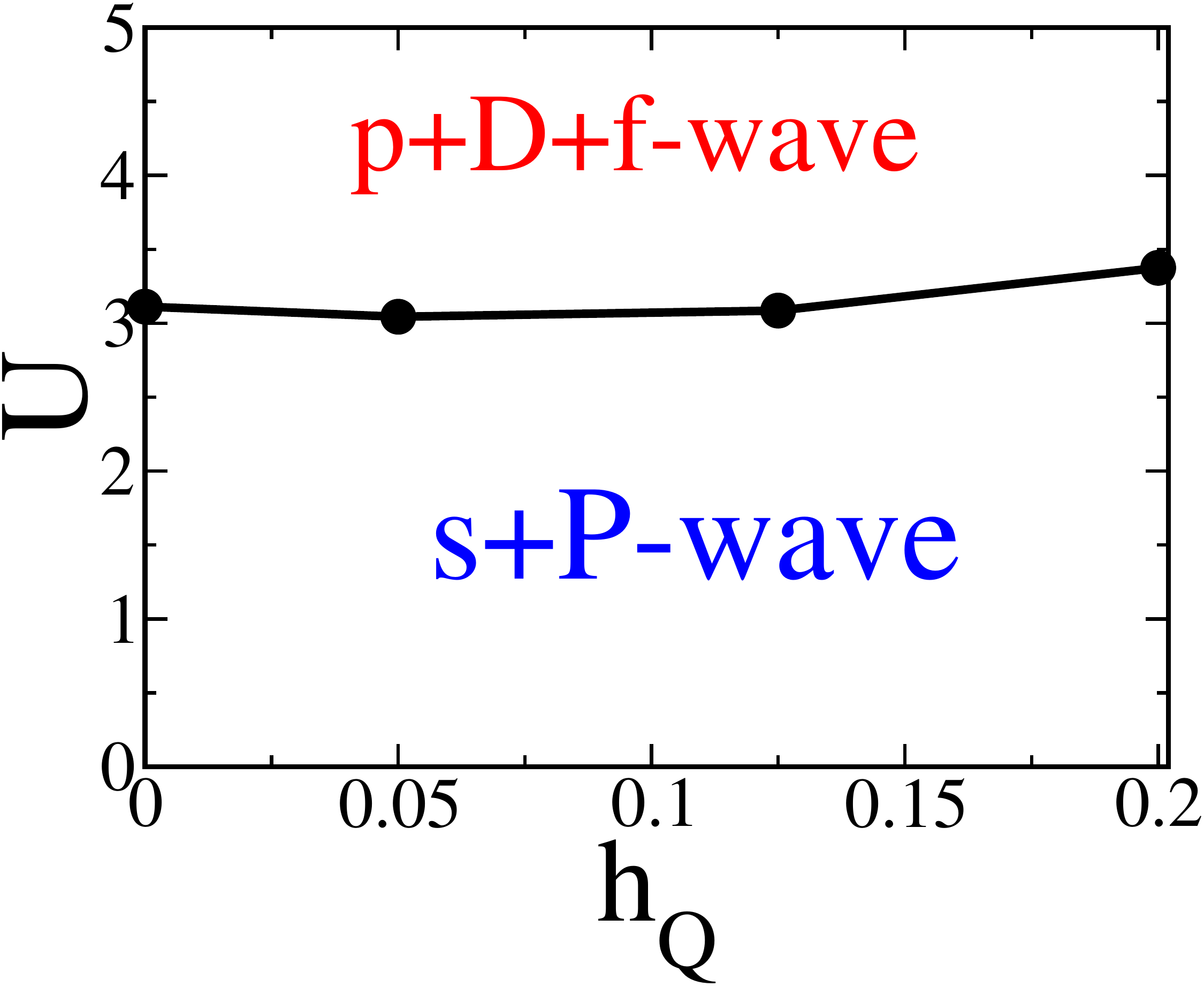}
\caption{(color online)
Phase diagram against $U$ and $h_{\rm Q}$ for $\alpha=0.3$. 
The $p$+$D$+$f$-wave state ($s$+$P$-wave state) is stable for a large (small) 
$U$. The solid line shows the boundary of two phases. 
 The \eli equation is solved in the $128 \times 128 \times 32$ lattice. 
}
    \label{fig:uhqphase}
  \end{center}
\end{figure}

 Next we discuss the stability of the SC state when the ASOC is introduced. 
 Figure 4 shows the $\alpha$-dependence of the eigenvalue of the \eli equation 
$\lambda$, for the $s$+$P$-wave and $p$+$D$+$f$-wave states. 
 The ASOC has two effects, namely, (i) the spin splitting
of the band and (ii) the pairing interaction. 
 The former is quantitatively important 
in most non-centrosymmetric superconductors where  
$|\alpha| \geq T_{\rm c}$.  
 It has been shown that the depairing effect due to (i) is minor 
for the spin singlet pairing state as well as for the spin triplet one 
with $\vec{d}(\k) \parallel \vec{g}(\k)$, while other 
spin triplet pairing states are destabilized.~\cite{rf:frigeri} 
 This is the reason why the $s$+$P$-wave state having the leading order 
parameter $\vec{d}(\k) \sim (-\sin k_{\rm y},\beta \sin k_{\rm x}, 0)$ is 
favored among $p$-wave states that have a six-fold degeneracy 
in the absence of the ASOC and AFM order. 
 Although the depairing effect due to the ASOC is almost avoided 
in this $s$+$P$-wave state, it does not vanish 
because the relation $\vec{d}(\k) \parallel \vec{g}(\k)$ is not strictly satisfied 
in the entire Brillouin zone. 
 It should be noted that the momentum dependence of the vector in the 
irreducible represantation of the point group is not unique. 
 Actually, the momentum dependences of the 
$d$-vector $\vec{d}(\k) \sim (-\sin k_{\rm y}, \sin k_{\rm x}, 0)$ and 
the $g$-vector $\vec{g}(\k)= (- v_{\rm y}(\k), v_{\rm x}(\k), 0) /\bar{v}$ 
are inequivalent although these vectors are in the same irreducible 
representation of the C$_{4v}$ point group. 
 Although the momentum dependence of the $d$-vector is assumed to be the 
same as that of the $g$-vector in many theories,~\cite{rf:fujimotoChi,
rf:samokhinChi,rf:frigeri,rf:frigeriChi,rf:mineevChi,rf:hayashiT1,
rf:fujimotoGap,rf:hayashiSD,rf:tadaRPA} 
this assumption is not supported by the microscopic theory since the 
momentum dependence of the $d$-vector is mainly determined by the 
pairing interaction. 
 Thus, Fig.~4 shows a steep decrease in $\lambda$ in the $s$+$P$-wave state 
when $\alpha$ is turned on. 
 This decrease arises from the depairing effect due to (i), whereby changes 
in the DOS due to band splitting is an additional source of the 
$ \alpha $-dependence.

\begin{figure}[htbp]
  \begin{center}
\includegraphics[width=7cm]{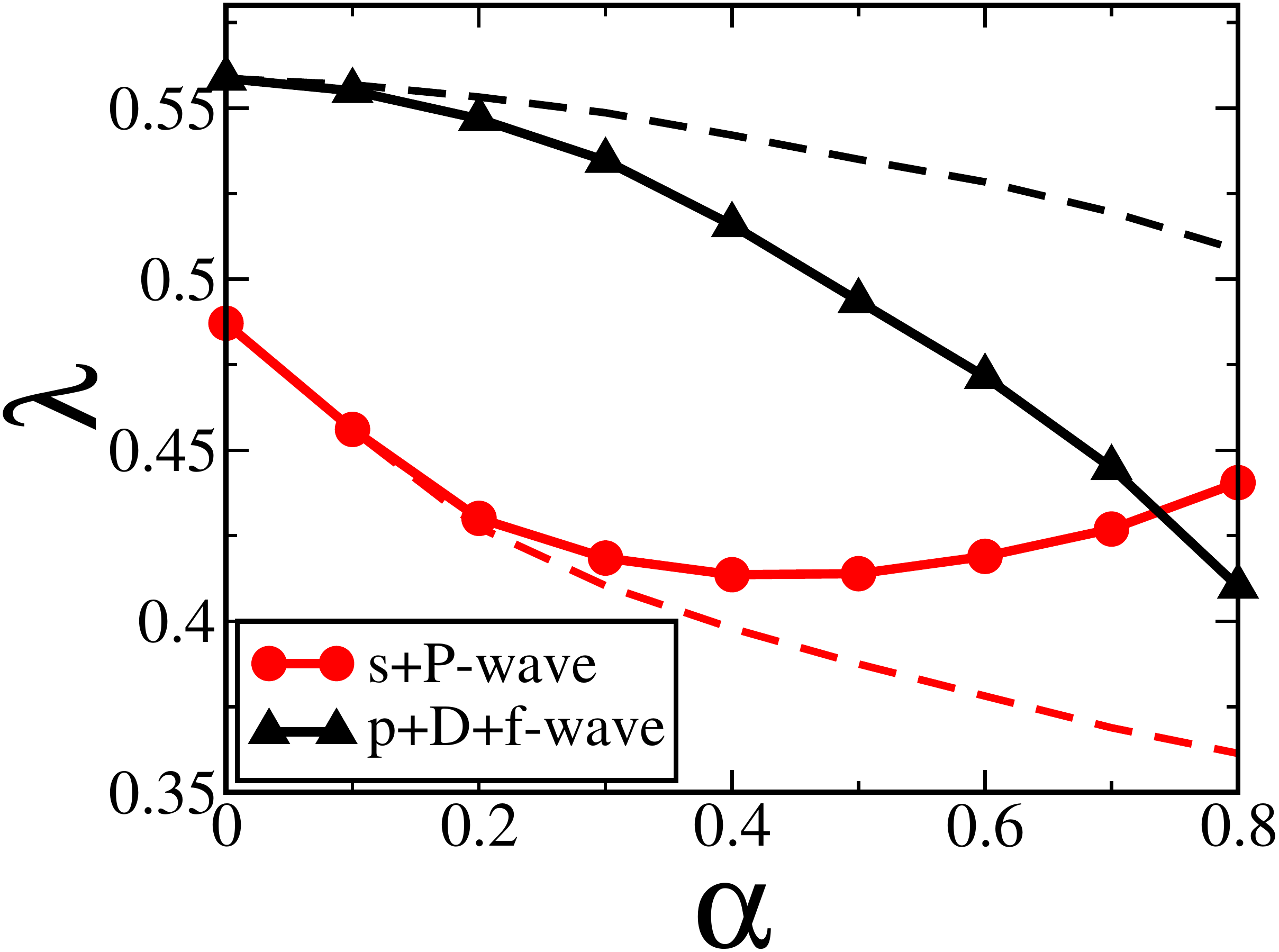}
\caption{(color online)
 Eigenvalues of \eli equation $\lambda$ for the $s$+$P$-wave (circles) and 
$p$+$D$+$f$-wave states (triangles). 
 We fix $U=4$ and $h_{\rm Q}=0$. 
 The \eli equation is solved in the $128 \times 128 \times 32$ lattice. 
 The dashed lines show $\lambda$ which is estimated by using the 
spin susceptibility  $\hat{\chi}_{i}(\q)$ for $\alpha=0$. 
}
    \label{fig:mzeropwave}
  \end{center}
\end{figure}

The effect (ii) of ASOC on the pairing interaction originates from the modification
of  the spin fluctuation. 
 This effect may be important 
in heavy Fermion systems since a large ASOC is likely induced through a  
strong $L$-$S$ coupling in $f$-orbitals. 
Figure~2(a) shows the suppression of the spin susceptibility at around 
$\q = (0,0,\pi)$, while that for other momenta is almost unchanged. 
 Since the spin fluctuations around $\q = (0,0,\pi)$ are the main source of 
the pairing interaction in the $p$+$D$+$f$-wave state, 
the eigenvalue $\lambda$ for the $p$+$D$+$f$-wave superconductivity 
monotonically decreases as 
$\lambda = \lambda(\alpha=0) - A \alpha^{2}/\e_{\rm F}^{2}$. 
In contrast, the eigenvalue $\lambda$ for the $s$+$P$-wave state 
shows a minimum and increases with increasing $\alpha$ for $\alpha > 0.4$. 
 Thus, the effect of ASOC on the pairing interaction favors the 
$s$+$P$-wave state rather than the $p$+$D$+$f$-wave state.

 These contrasting effects of ASOC arise from the anisotropy of 
helical spin fluctuation. To clarify this point we show the 
eigenvalue of \eli equation $\lambda$ (dashed lines) in which 
the spin susceptibility $\hat{\chi}_{i}(\q)$ is estimated for $\alpha=0$. 
 We see that the $\lambda$ is increased for the $s$+$P$-wave state 
by the modification of spin susceptibility due to ASOC. 
 Thus, the $s$+$P$-wave state is favored by the modified spin fluctuation 
although the magnitude of spin fluctuation is suppressed 
at around $\q = (0,0,\pi)$ by the ASOC. 
 These results indicate that the anisotropy of helical spin fluctuation 
enhances the $s$+$P$-wave state.

 We confirmed that the subdominant component of the order parameter 
($s$-wave component in the $s$+$P$-wave state, 
$p$- and $f$-wave components in the $p$+$D$+$f$-wave state) 
 grows almost linearly with increasing $\alpha$, 
and that the momentum dependence of each component 
is almost independent of $\alpha$.

 The eigenvalue of the \eli equation is decreased by the AFM order 
owing to the loss of quasiparticle DOS. 
 Therefore, the superconductivity is suppressed by the AFM order 
independently of the pairing symmetry. 
 The relative stability of the $s$+$P$-wave and $p$+$D$+$f$-wave states 
is hardly affected by the AFM order, as shown in Fig.~3. 
 The stability of the interlayer $d$-wave state against the 
A-type AFM order has been claimed~\cite{rf:shimaharadwave} 
by assuming the quasi-two-dimensional Fermi surface. 
 However, this is not the case in our model that assumes the 
three-dimensional $\beta$-band. 
 The \Tc of \Pt decreases if the AFM order decreases upon 
the application of pressure~\cite{rf:yasuda,rf:takeuchiP,rf:tateiwa} and 
seems to be incompatible with our result. 
 However, this pressure dependence may be due to the suppression of 
electron correlation by increasing pressure.

 It is expected that the AFM order leads to much more significant depairing 
effects on the intralayer $d$-wave and interlayer $p$-wave states 
because these Cooper pairings are directly broken by the A-type AFM order. 
 The stability of the SC state against the AFM order~\cite{rf:yasuda,
rf:tateiwa,rf:takeuchiP}  implies 
the interlayer $p$+$D$+$f$-wave or intralayer  $s$+$P$-wave state in \Pt 
that is identified in our calculation.

\subsection{Superconducting gap}

 We investigate here the gap structure of both the $s$+$P$-wave and 
$p$+$D$+$f$-wave states and discuss the consistency with the line node 
behavior observed in \Pt at ambient 
pressure.~\cite{rf:bonalde,rf:izawa,rf:takeuchiC,rf:mukudaT1} 

 The quasiparticle spectrum in the SC state is 
obtained by diagonalizing the 8 $\times$ 8 matrix using 
\begin{eqnarray}
\label{eq:8-8-matrix}
\hat{H}_{\rm s}(\k)
=
\left(
\begin{array}{cc}
\hat{H}(\k) & -\Delta_{0} \hat{\Delta}(\k) \\
-\Delta_{0} \hat{\Delta}^{\dag}(\k)  & -\hat{H}(-\k)^{\rm T} \\
\end{array}
\right), 
\end{eqnarray}
where $\hat{\Delta}(\k)$ is the SC order parameter in the spin basis expresses as 
\begin{eqnarray}
\label{eq:Delta-matrix}
\hat{\Delta}(\k) 
= 
\left(
\begin{array}{cc}
\Delta_{1,s,s'}(\k) & \Delta_{2,s,s'}(\k) \\
\Delta_{2,s,s'}(\k+\Q) & \Delta_{1,s,s'}(\k+\Q) \\
\end{array}
\right). 
\end{eqnarray}
 In the following calculations the matrix element of $\hat{\Delta}(\k)$ 
is determined from the linearized \eli equation by assuming that 
the momentum and spin dependences of the order parameter are weakly 
dependent on temperature for $T \leq T_{\rm c}$. 
 We solve the \eli equation at $T = 0.02 > T_{\rm c}$, having 
confirmed that the matrix $\hat{\Delta}(\k)$ is 
almost independent of temperature for $T < 0.1$. 
 The same assumption has been adopted in other studies of 
multi-orbital superconductivity.~\cite{rf:nomura,rf:yanasecobaltSO2}

 Since the amplitude of the SC order parameter $\Delta_{0}$ is arbitrary 
in the linearized \eli equation, we here choose $\Delta_{0}$ so that 
the magnitude of the maximal gap is $\Delta_{\rm g}=0.1$ in our energy units. 
 Although this magnitude may be large compared with the 
energy scale $ \alpha $ or $ h_{\rm Q} $, we adopt this value for 
numerical accuracy, having confirmed that the lower values 
of $ \Delta_{\rm g} $ do not alter the result qualitatively. 
 We define the quasiparticle DOS $\rho(\e)$ as 
$\rho(\e) = \frac{1}{4N}\sum_{i}\sum'_{\k} \delta(\e-E_{i}(\k))$ 
where $\sum'_{k}$ denotes the summation within the range $|k_{\rm z}| < \frac{\pi}{2}$. 
 The eigenvalues $E_{8}(\k) > E_{7}(\k) > .....  > E_{1}(\k)$ satisfy the 
relation $E_{i}(\k)=-E_{9-i}(\k)$.

 It is more transparent to describe the SC order parameter 
in the band basis, which is obtained by unitary transformation using
\begin{eqnarray}
\label{eq:Delta-unitary}
\hat{\Delta}_{\rm band}(\k) 
= 
\hat{U}^{\dag}(\k) \hat{\Delta}(\k) \hat{U}^{*}(-\k). 
\end{eqnarray}
 The unitary matrix $\hat{U}(\k)$ diagonalizes the 
unperturbed Hamiltonian as 
\begin{eqnarray}
\label{eq:4-unitary}
\hat{U}^{\dag}(\k) \hat{H}(\k) \hat{U}(\k) = (\ee_{i}(\k) \delta_{ij}). 
\end{eqnarray}
 The SC gap in the $\gamma$-th band is obtained as 
\begin{eqnarray}
\label{eq:band-basis-gap}
&& \hspace*{-15mm} 
\Delta_{\gamma}(\k)=\Delta_{0} \Psi_{\gamma}(\k), 
\end{eqnarray}
where $\Psi_{\gamma}(\k)$ is the $(\gamma\gamma)$ component of the 
matrix $\hat{\Delta}_{\rm band}(\k)$. 
%
%
%
 Since the relation $T_{\rm c} \ll |\alpha|$ is satisfied in most of 
the non-centrosymmetric superconductors, the relation 
$|\Delta_{\rm band}^{ij}(\k)| \ll 
|\alpha \hat{g}(\k)|,{\rm max}\{h_{\rm Q},|\e(\k)-\e(\k+\Q)|\}$
is valid for each (ij) component of $\hat{\Delta}_{\rm band}(\k)$ 
except for the special momentum such as $\k=(0,0,k_{\rm z})$. 
 Therefore, the off-diagonal components of $\hat{\Delta}_{\rm band}(\k)$ 
hardly affect the electronic state, and 
the quasiparticle excitations $E_{i}(\k)$ are approximated to 
$\pm E_{\gamma}^{\rm band}(\k)$ with 
$E_{\gamma}^{\rm band}(\k)^{2} = 
\ee_{\gamma}(\k)^{2} + |\Delta_{\gamma}(\k)|^{2}$. 
 Thus, the SC gap in the $\gamma$-th band is described by 
$|\Delta_{\gamma}(\k)|$.

 It is clear that the $p$+$D$+$f$-wave state has a horizontal line node 
protected by the symmetry because all of the matrix elements of 
$\hat{\Delta}(\k)$ are zero at $k_{\rm z}=0$.
 This is consistent with the 
experiments in \Ptf.~\cite{rf:bonalde,rf:izawa,rf:takeuchiC,rf:mukudaT1} 
 The coefficient $c_1$ of the linear term in the DOS 
($\rho(\e) =c_1 \e$) increases in the AFM state because 
the pairing state changes from the chiral 
$d_{\rm xz} \pm {\rm i} d_{\rm yz}$-wave state in the paramagnetic state 
to the $d_{\rm xz}$-wave state in the AFM state (see Fig.~3 of ref.~28).

\begin{figure}[htbp]
  \begin{center}
\includegraphics[width=6cm]{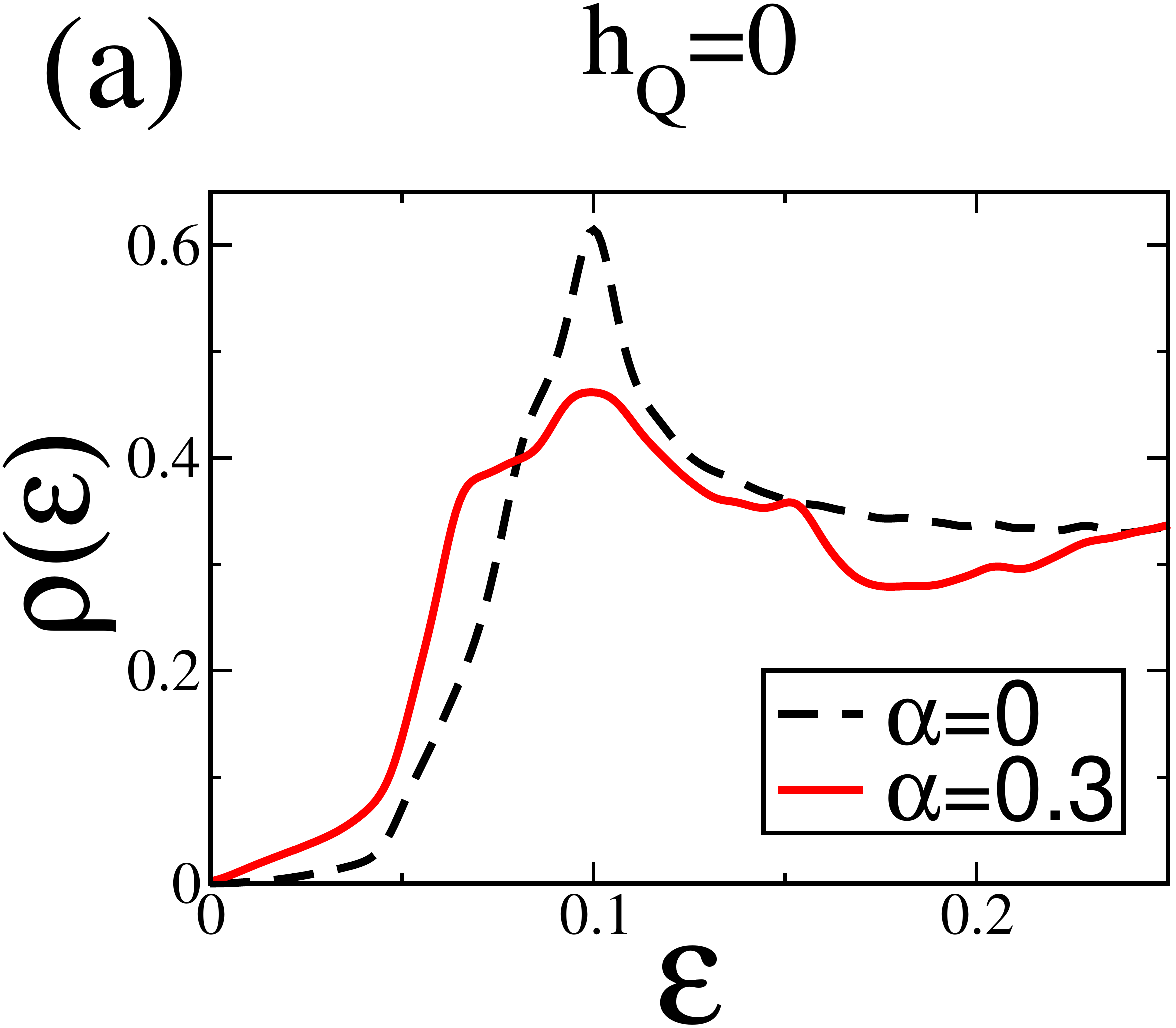}
\\
\vspace*{5mm}
\includegraphics[width=6cm]{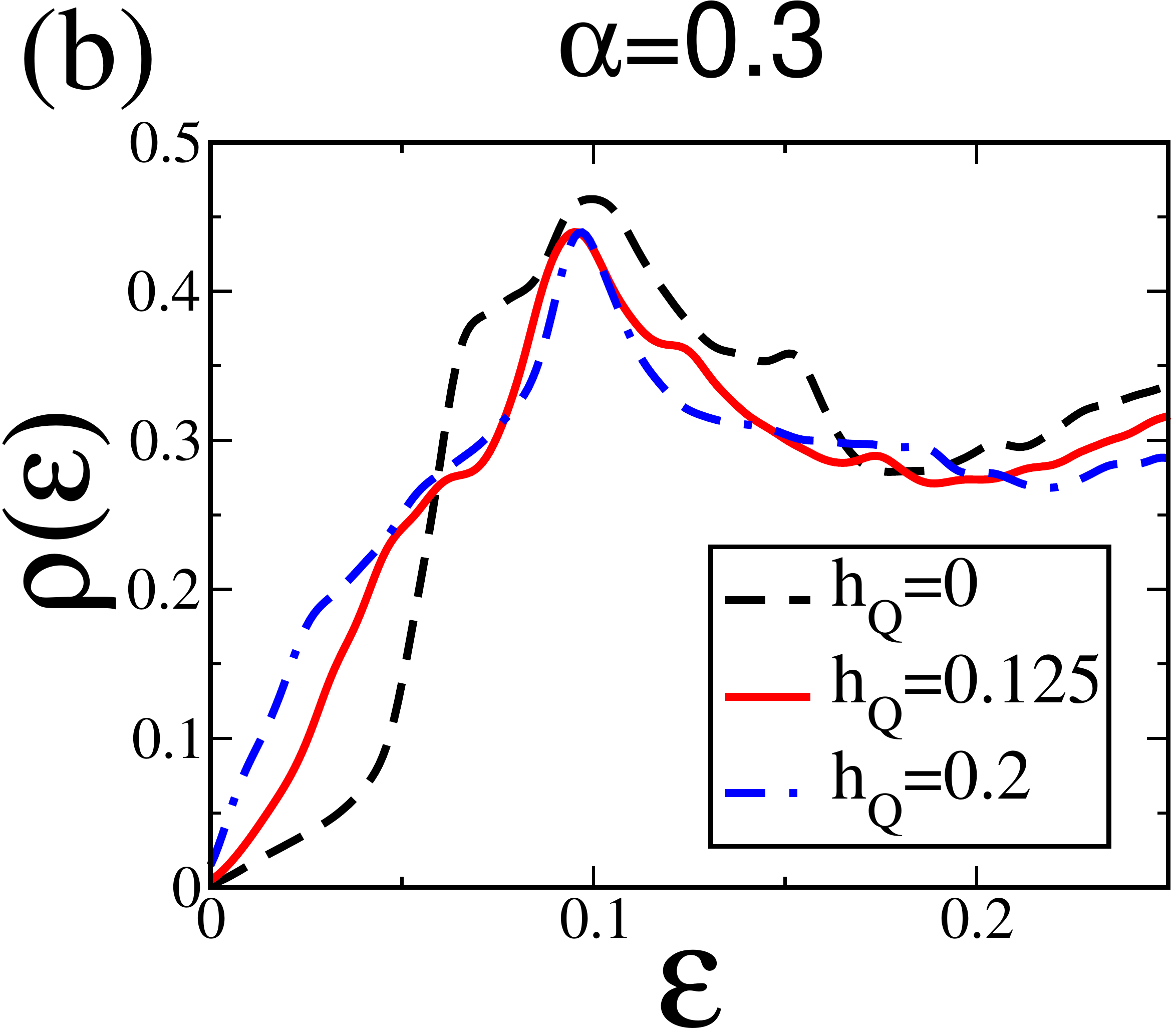}
\caption{(color online)
DOS $\rho(\e)$ in the $s$+$P$-wave state at $U=4$. 
(a) Paramagnetic state for $\alpha=0$ (dashed line) and $0.3$ (solid line). 
(b) AFM state with ASOC $\alpha=0.3$ for $h_{\rm Q}=0.125$ 
(solid line) and $h_{\rm Q}=0.2$ (dash-dotted line). 
 We show the results for $\e > 0$ because $\rho(\e)$ is particle-hole 
symmetric owing to its definition. 
 The \eli equation is solved in the $64 \times 64 \times 32$ lattice and 
the DOS is calculated in the $384 \times 384 \times 384$ lattice. 
}
    \label{fig:dos}
  \end{center}
\end{figure}

\begin{figure}[htbp]
  \begin{center}
\vspace*{-1.5cm}
\hspace*{-2cm}
\includegraphics[width=12cm]{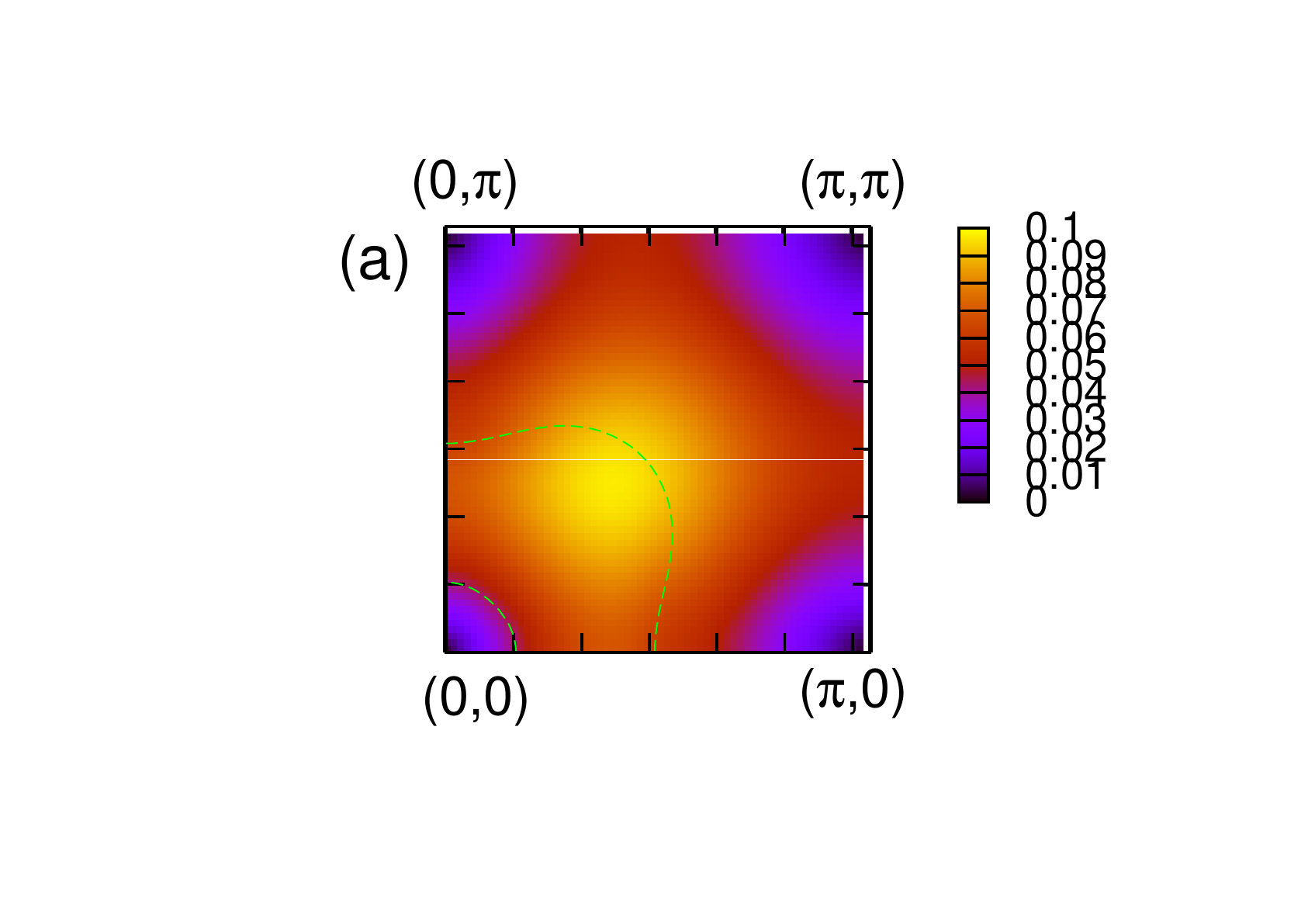}
\\
\vspace*{-2.5cm}
\hspace*{-2cm}
\includegraphics[width=12cm]{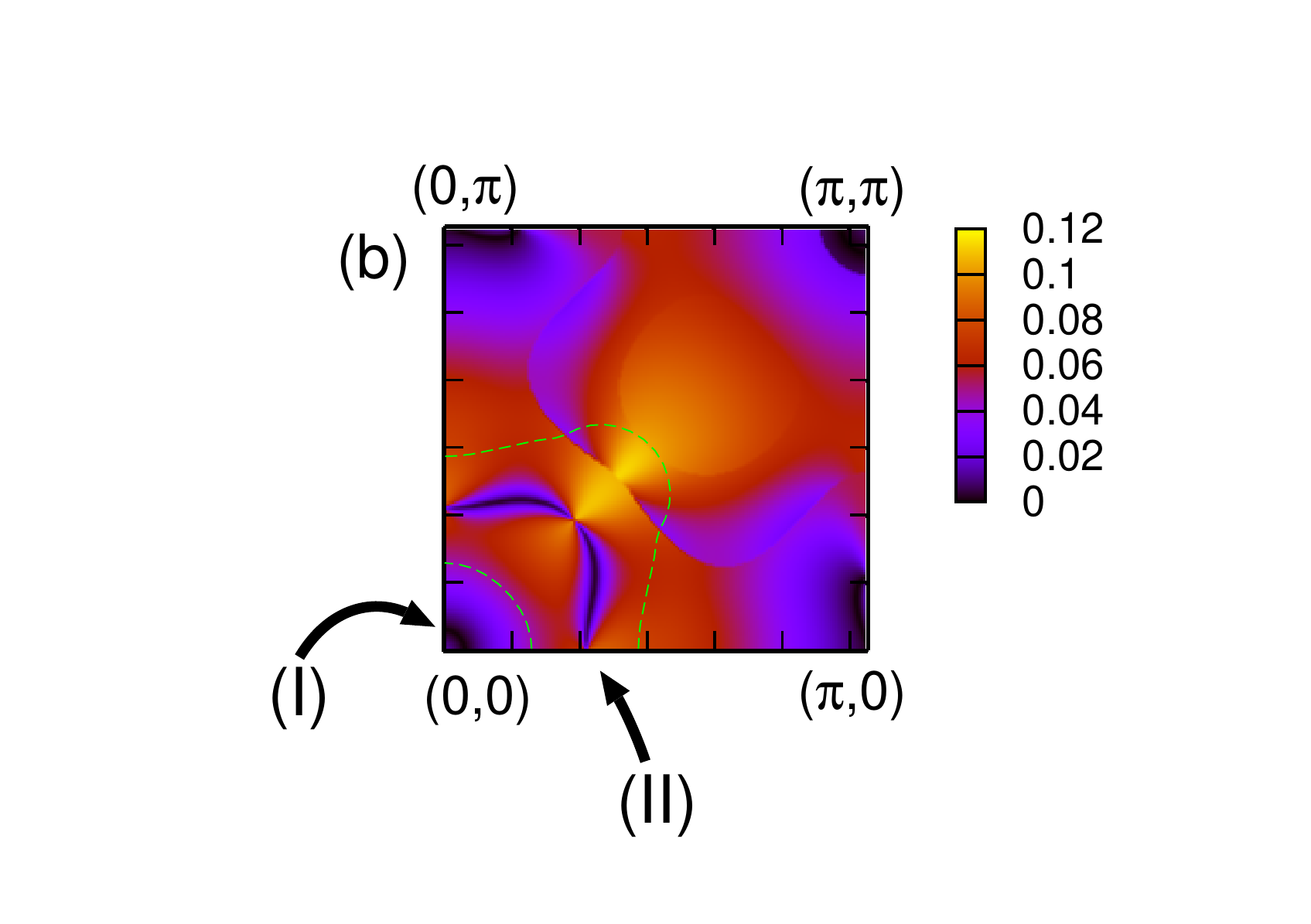}
\\
\vspace*{-1.5cm}
\hspace*{-2cm}
\includegraphics[width=12cm]{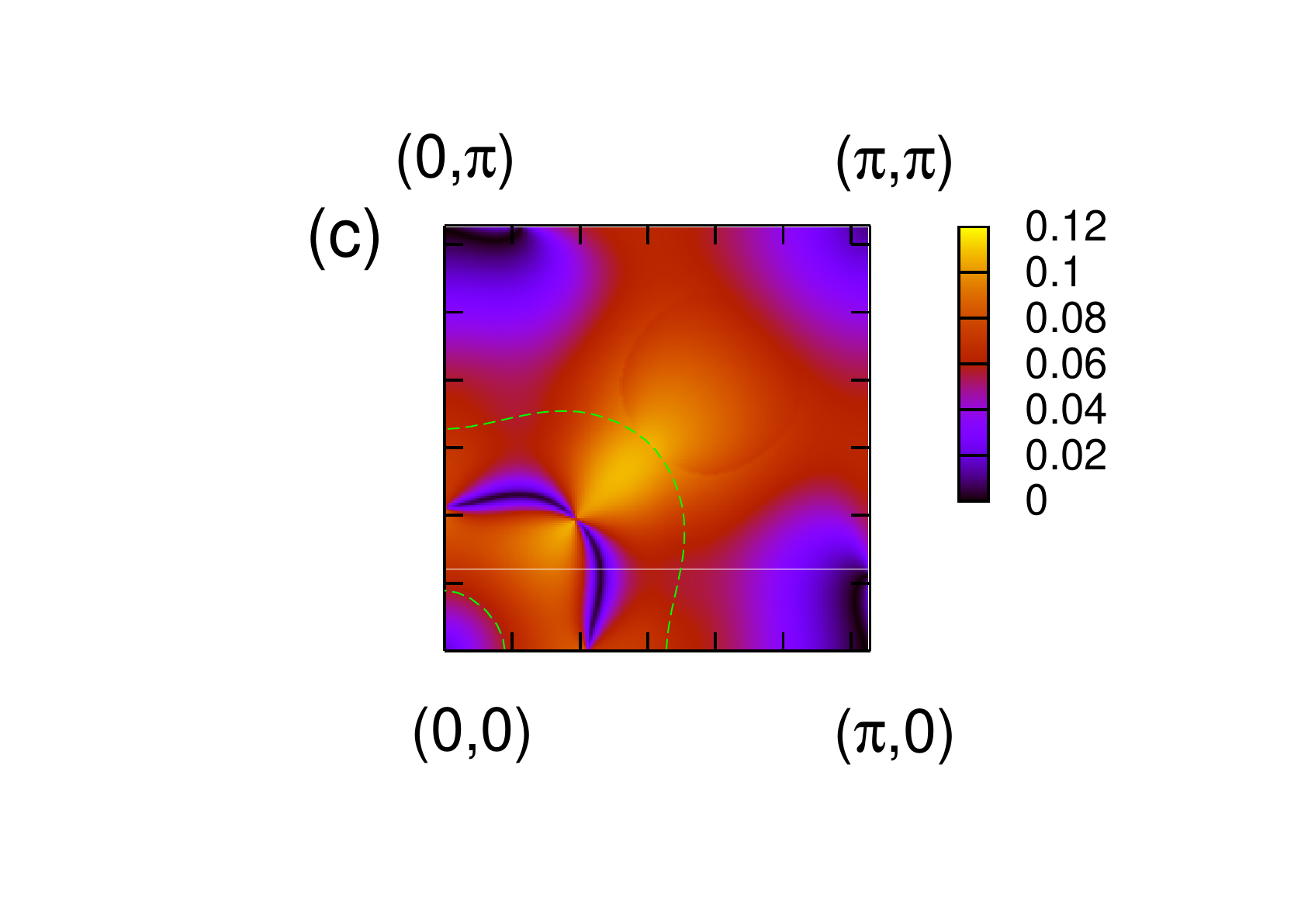}
\\
\vspace*{-1cm}
\caption{(color online)
SC gap in the $s$+$P$-wave state at $k_{\rm z}=\frac{\pi}{3}$ 
in the paramagnetic state ($h_{\rm Q}=0$).  
(a) SC gap in the absence of the ASOC ($\alpha=0$). 
We show the SC gaps $|\Delta_{3}(\k)|$ and $|\Delta_{4}(\k)|$ 
for $\alpha=0.3$ in (b) and (c), respectively. 
 The SC gaps $|\Delta_{1}(\k)|$ and $|\Delta_{2}(\k)|$ 
are not shown because the quasiparticle DOS is small in these bands. 
 The positions of the line nodes arising from the mechanisms (I) and (II) 
are shown by the arrows in (b) (see the text). 
 The thin dashed lines show the Fermi surface at $k_{\rm z}=\frac{\pi}{3}$. 
 Although the zeros of SC gap do not intersect with the Fermi surface 
at $k_{\rm z}=\frac{\pi}{3}$, the line nodes exist on the Fermi surface 
at another $k_{\rm z}$. Note that the $\beta$-band has a three-dimensional 
Fermi surface, while the SC gap in the $s$+$P$-wave state is nearly independent 
of $k_{\rm z}$. 
}
\label{fig:SCgap}
  \end{center}
\end{figure}

 We investigate here the accidental line node of the SC gap 
in the $s$+$P$-wave state in detail. 
 The quasiparticle DOS $\rho(\e)$ is shown in Fig.~5 and 
the SC gaps $|\Delta_{\gamma}(\k)|$ for $\gamma=3,4$ 
are shown in Figs.~6 and 7, respectively. 
The paramagnetic state is assumed in Fig.~6,  
while the AFM state is assumed in Fig.~7.

 The SC gap in the absence of the ASOC and AFM order (Fig.~6(a)) is 
approximated to be 
$|\Delta(\k)| \sim \sqrt{\sin k_{\rm x}^{2}+\sin k_{\rm y}^{2}}$, 
which has two point nodes in the [001] direction in contrast to the 
experimental results.~\cite{rf:bonalde,rf:izawa,rf:takeuchiC,rf:mukudaT1} 
 The DOS at low energies is quadratic as shown by $\rho(\e) \sim c_{2} \e^{2}$ 
and the coefficient $c_{2}$ is small owing to the small DOS in the [001] 
direction (dashed line in Fig.~5(a)).

 The line nodes are induced by the ASOC through the 
following two mechanisms. 

\noindent
(I) Admixture with an $s$-wave order parameter.

\noindent
(II) Mismatch of the $d$-vector and $g$-vector.

 The first one (I) has been proposed by Frigeri 
{\it et al.}~\cite{rf:frigeriImp} 
and its contributions to the NMR $1/T_{1}T$ and superfluid density 
have been investigated by Hayashi {\it et al.}~\cite{rf:hayashiT1,rf:hayashiSD} 
 In the absence of the AFM order, the SC gap is expressed as 
$\pm \Phi(\k) + \d(\k) \cdot \tilde{g}(\k)$ with 
$\tilde{g}(\k)=\vec{g}(\k)/|\vec{g}(\k)|$. 
 The $s$-wave and $p$-wave order parameters 
are approximated to be $\Phi(\k) \sim \Phi(\vec{0},k_{\rm z})$ and 
$\d(\k) \cdot \tilde{g}(\k) \sim c(k_{\rm z}) |\k_{\parallel}|$ 
at around $\k_{\parallel}=(k_{\rm x},k_{\rm y})=(0,0)$, respectively. 
 Therefore, the SC gap vanishes on the line 
$ |\k_{\parallel}| = |\Phi(\vec{0},k_{\rm z})|/c(k_{\rm z})$ 
in half of the bands, while the other bands have a full gap. 
 We show the SC gaps $|\Delta_{3}(\k)|$ and $|\Delta_{4}(\k)|$ 
at $k_{\rm z}=\pi/3$ in Figs.~6(b) and 6(c), respectively. 
 The line node actually appears in $|\Delta_{3}(\k)|$ in the vicinity of 
$\k_{\parallel}=(0,0)$ (shown by the arrow (I)). 
 However, the line node arising from the mechanism (I) induces only a tiny 
linear term $\rho(\e) \sim c_{1} \e$ with $c_{1} \propto |\alpha|$ 
because the length of the line node is very small, as shown in Fig.~6(b).

 We find another line node arising from the mechanism (II) 
at around $|k_{\parallel}| = \pi/3$ 
(see the arrow (II) in Fig.~6(b)). 
 This line node originates from the topological character of the $g$-vector. 
 According to the assumptions 
$g_{\rm x}(\k)= - v_{\rm y}(\k)/\bar{v}$ and 
$g_{\rm y}(\k)=   v_{\rm x}(\k)/\bar{v}$, 
the $g$-vector has a singularity not only on the [001] line but also 
on the line at around $(k_{\rm x},k_{\rm y})=(0.4\pi,0.4\pi)$. 
 The $g$-vector rotates around the singular point, and therefore 
the relation $\vec{d}(\k) \perp \vec{g}(\k)$ is satisfied on a line. 
 The SC gap $\pm \Phi(\k) + \d(\k) \cdot \tilde{g}(\k)$ 
vanishes around this line because the $s$-wave component $|\Phi(\k)|$ is 
much smaller than the $p$-wave component $|\d(\k)|$. 
 This is a general mechanism for the line node in 
the non-centrosymmetric superconductor predominated by the spin triplet 
pairing. 
 However, it is not clear whether this line node exists in CePt$_3$Si 
because it depends on the detailed momentum dependence of the $g$-vector. 
 For example, this line node does not appear if we assume 
$\vec{g}(\k)=(-\sin k_{\rm y},\sin k_{\rm x},0)$. 
 Anyway, the low-energy excitation arising from the ASOC 
is small because of the steep increase in SC gap around the 
line node, as shown in the schematic figure (Fig.~8). 
 Figure~5(a) actually shows a small coefficient $c_1$ of 
the linear term $\rho(\e) =c_1 \e$ (solid line). 
 This linear term mainly arises from the line node induced by 
the mechanism (II) and the contribution of the line node (I) 
is negligible.

\begin{figure}[htbp]
  \begin{center}
\vspace*{-1cm}
\hspace*{-25mm}
\includegraphics[width=12cm]{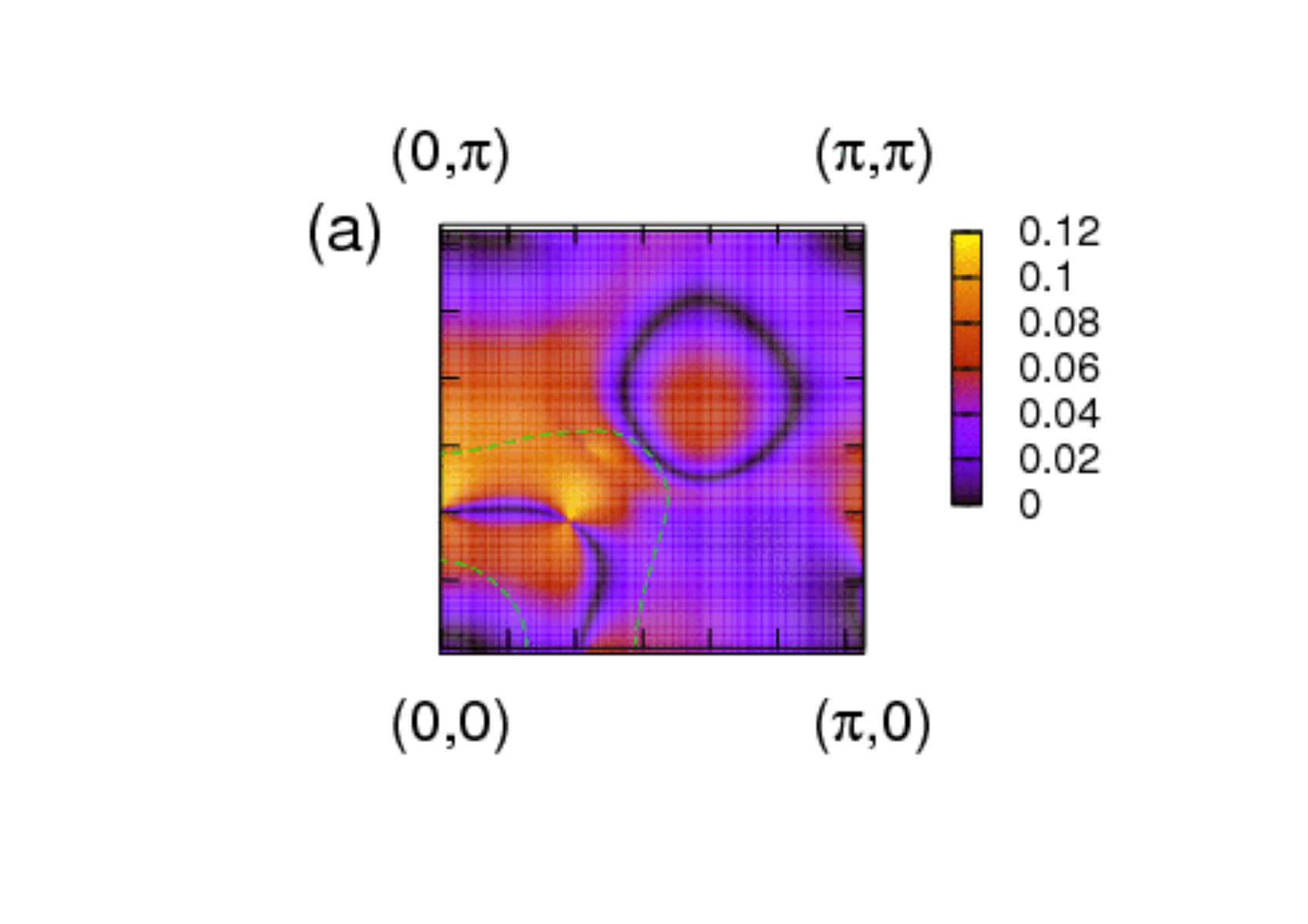}
\\
\vspace*{-1.5cm}
\hspace*{-25mm}
\includegraphics[width=12cm]{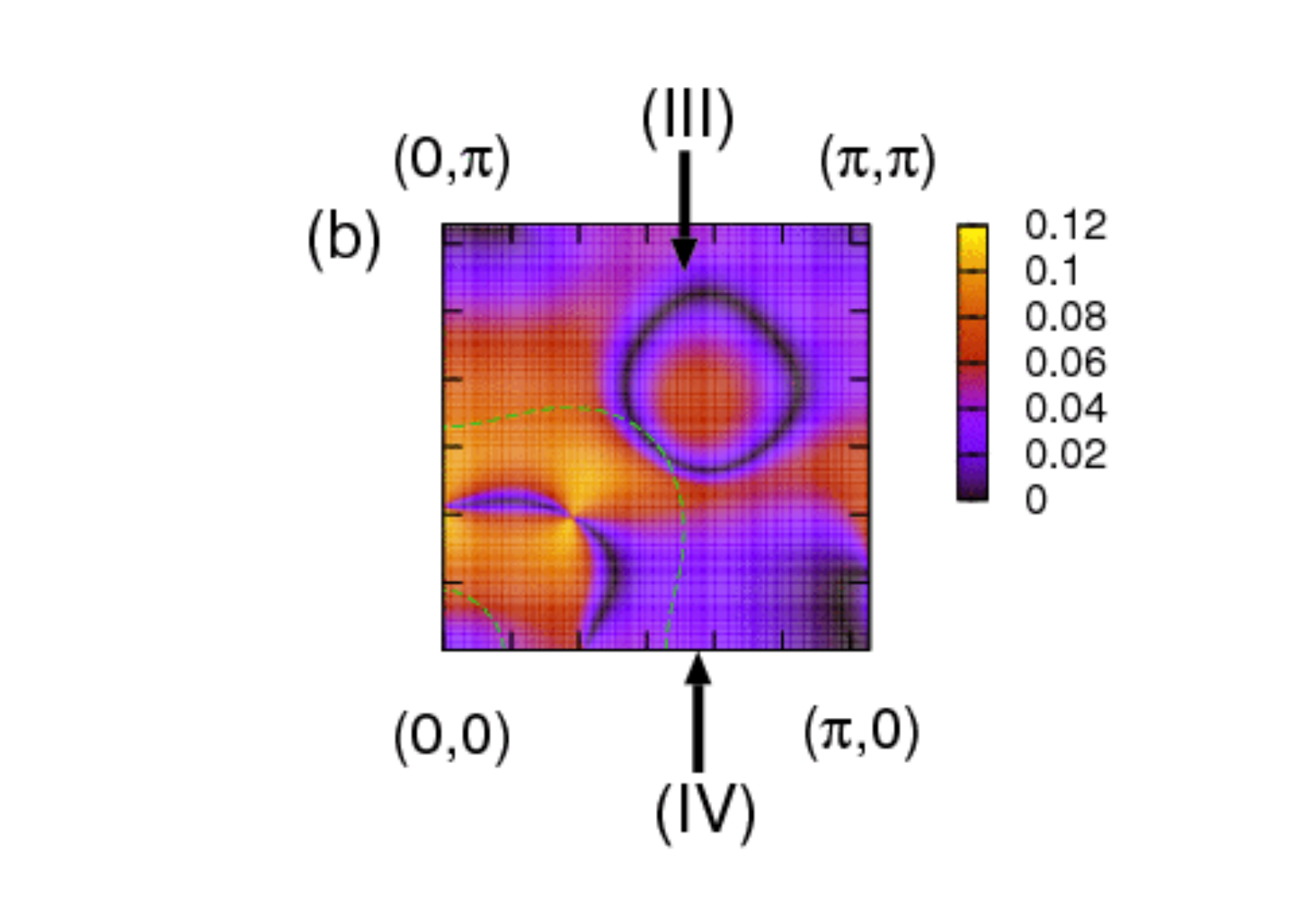}
\\
\vspace*{-0.5cm}
\caption{(color online)
SC gap in the $s$+$P$-wave state with the AFM order. 
We show (a) $|\Delta_{3}(\k)|$ and (b) $|\Delta_{4}(\k)|$ 
at $k_{\rm z}=\frac{\pi}{3}$ for $\alpha=0.3$ and $h_{\rm Q}=0.125$.  
 The positions of the line nodes arising from the mechanism 
(III) and (IV) are shown by the arrows in (b). 
}
  \end{center}
\end{figure}

\begin{figure}[ht]
\begin{center}
\includegraphics[width=4.5cm]{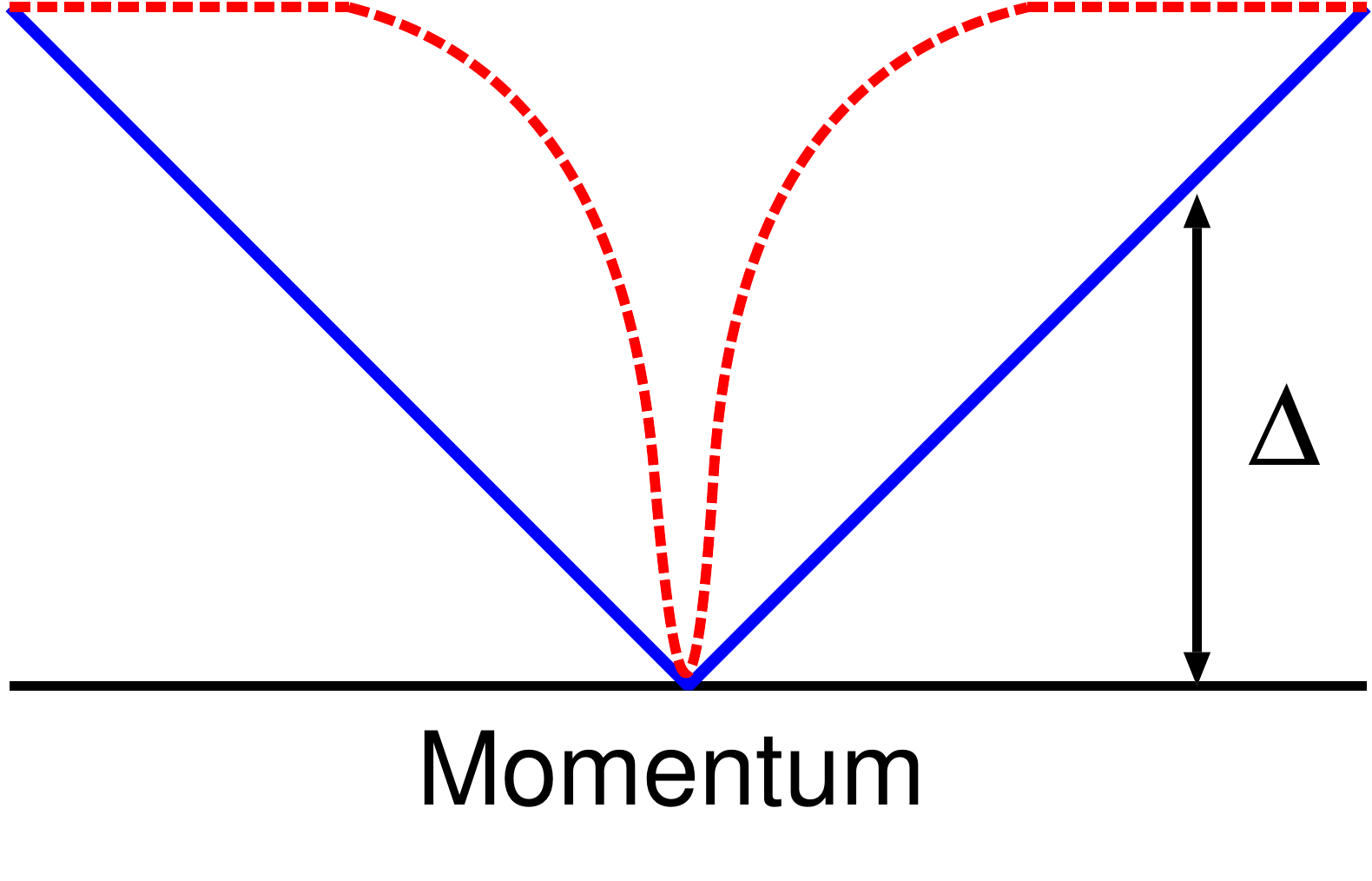}
\caption{(color online)
Schematic figure showing the momentum dependence of SC gap 
along the Fermi surface. 
The solid line shows the usual line node protected by the 
symmetry (for example, the $d_{\rm x^{2}-y^{2}}$-wave superconductor). 
 The accidental line nodes (II) in Fig.~6(b) and (III) in Fig.~7(b) 
show a steep increase in the SC gap around the gap node (dashed line). 
} 
\end{center}
\end{figure}

 The DOS at low energies is markedly increased by the AFM order 
owing to the following two effects. 

\noindent
 (III) Folding of the Brillouin zone. 

\noindent
 (IV) Mixing of the $p$-wave order parameter 
between the leading part 
$\vec{d}(\k) \sim (-\sin k_{\rm y},\sin k_{\rm x},0)$ and the 
admixed part $\hat{d}(\k)=(\sin k_{\rm y},\sin k_{\rm x},0)$.

 The former arises from the pair-breaking effect due to the band mixing,  
which has been investigated by Fujimoto.~\cite{rf:fujimotoGap} 
 In contrast to ref.~31, the line node appears not only at 
$k_{\rm z}=\pi/2$ but also at around $(k_{\rm x},k_{\rm y})=(\pi/2,\pi/2)$ 
(see Fig.~7(b)) in our case because of the band structure of the $\beta$-band. 
 However, the DOS arising from (III) is not quantitatively important 
when $h_{\rm Q} \ll W$ because of the steep increase in the SC gap around 
the line node, as shown in Fig.~8.

 Actually, the low-energy excitations in the $s$+$P$-wave state 
are mainly induced by the effect (IV). 
 The {\it a}- and {\it b}-axes in the tetragonal lattice 
are no longer equivalent in the presence of the AFM order. 
 Therefore, the $p$-wave order parameter is modified to 
$\vec{d}(\k) = (-\sin k_{\rm y}, \beta \sin k_{\rm x},0)$ with 
$\beta \ne 1$. 
 This change can be viewed as the mixing of the leading part 
$\vec{d}(\k) = (-\sin k_{\rm y},\sin k_{\rm x},0)$ with the 
admixed part $\hat{d}(\k)=(\sin k_{\rm y},\sin k_{\rm x},0)$, which leads 
to the rotation of the $d$-vector. 
 According to the result obtained using the RPA theory, $\beta$ decreases 
with increasing $h_{\rm Q}$.  
 Then, many low-energy excitations are induced at around 
$k_{\rm y}=\pi/6$, as shown in Fig.~7(b). 
 The SC gap in the $4$-th band (Fig.~7(b)) is further decreased 
at around $k_{\rm y}=\pi/6$ by the admixture with an $s$-wave order parameter. 
 The DOS clearly shows a linear dependence in Fig.~5(b), 
which is consistent with the experimental results in \Pt 
at ambient pressure.~\cite{rf:bonalde,rf:izawa,rf:takeuchiC,rf:mukudaT1} 
 We have shown that the rotation of the $d$-vector is also the main source of 
the anomalous paramagnetic properties of \Ptf.~\cite{rf:yanasehelical}

\subsection{Specific heat and NMR 1/T$_1$T}

 The pressure dependence of the SC state is a decisive test 
for validating the theory of \Pt as well as of \Rh and \Irf. 
 According to the experimental result of  
\Ptf,~\cite{rf:tateiwa,rf:yasuda,rf:takeuchiP} 
the AFM order is suppressed at a pressure $P \sim 0.6$GPa,  
although the superconductivity survives at high pressures $P > 0.6$GPa. 
 Therefore, the role of the AFM order can be studied experimentally 
by measuring the pressure dependence of the SC state.  
 If the $s$+$P$-wave state is realized in \Pt and the AFM order is the 
main source of line nodes, the number of low-energy  excitations decreases 
under pressure.  
 This theoretical result can be tested by measuring the pressure dependence of  
specific heat, NMR $1/T_{1}T$, superfluid density, thermal conductivity, 
and other quantities. 
 We now calculate specific heat and NMR $1/T_{1}T$ for a 
future experimental test.

 To discuss these quantities, we adopt the same assumption in \S4.2. 
 We here calculate the amplitude of the SC gap, $\Delta_{0}$, in eq.~(19) 
by solving the gap equation  
\begin{eqnarray}
\label{eq:gap-equation}
&& \hspace*{-10mm}  
1 = g \sum'_{k} |\Psi_{\gamma}(\k)|^{2} 
\tanh \frac{E_{\gamma}^{\rm band}(\k)}{2 T}/2 E_{\gamma}^{\rm band}(\k), 
\end{eqnarray}
which is obtained as a mean field solution of the effective 
model in the band basis given as
\begin{eqnarray}
\label{eq:effective-model}
%
&& \hspace*{-10mm}  
H = \sum'_{k}\sum_{\gamma=1}^{4} 
\ee_{\gamma}(\k) d_{\k,\gamma}^{\dag} d_{\k,\gamma} 
- \frac{1}{2} g \sum'_{k,k'} b_{\k}^{\dag} b_{\kk},  
\\ && \hspace*{-10mm}   
b_{\k}^{\dag} = \sum_{\gamma} \Psi_{\gamma}(\k) 
d_{\k,\gamma}^{\dag} d_{-\k,\gamma}^{\dag}. 
\end{eqnarray}
 The SC order parameter obtained in the linearized \eli equation 
(eqs.~(5)-(7)) is reproduced using this model. 
 We choose $g$ so as to obtain $T_{\rm c}=0.05$.  
 We have confirmed that the smaller $g$ and \Tc do not 
qualitatively alter the following results.

 The quasiparticle excitation $E_{i}(\k)$ is determined 
using eq.~(19) with $\Delta_{0}$ determined using eq.~(24). 
 The Sommerfeld coefficient $C/T$ is obtained as 
\begin{eqnarray}
\label{eq:specific-heat}
&& \hspace*{-10mm}  
C/T = \frac{\partial S}{\partial T}, 
\\ && \hspace*{-10mm} 
S = - \sum'_{k} \sum_{i=1}^{8} [f_{i,\k} \log f_{i,\k} + 
(1-f_{i,\k}) \log (1-f_{i,\k})],  
\nonumber \\
\end{eqnarray}
where $f_{i,\k}$ is the Fermi distribution function 
$f_{i,\k}=(1+\exp (E_{i}(\k)/T))^{-1}$.

 We calculate NMR $1/T_{1}T$ as 
\begin{eqnarray}
\label{eq:NMR-T1T}
&& \hspace*{-10mm}  
1/T_{1}T = {\rm Im} \chi_{\rm L}(\Omega)/\Omega|_{\Omega \rightarrow 0}, 
\\ && \hspace*{-10mm}  
\chi_{\rm L}({\rm i}\Omega_{n}) = 
- \sum_{k,k',\omega_{n}} \sum_{i=1}^{2}
[
G_{i,\uparrow\uparrow}(\kk,{\rm i}\omega_{n} + {\rm i}\Omega_{n}) 
G_{i,\downarrow\downarrow}(\k,{\rm i}\omega_{n}) 
\nonumber \\ && \hspace*{15mm}  
- 
F^{\dag}_{i,\downarrow\uparrow}(\kk,{\rm i}\omega_{n} + {\rm i}\Omega_{n})
F_{i,\downarrow\uparrow}(\k,{\rm i}\omega_{n}) 
], 
\end{eqnarray}
where $\hat{G}_{i}(\k,{\rm i}\omega_{n})$ and 
$\hat{F}_{i}(\k,{\rm i}\omega_{n})$ are the normal and anomalous 
Green functions in the SC state, respectively. 
 We ignore the momentum dependence of the hyperfine coupling 
constant and the exchange enhancement due to the electron correlation 
for simplicity. 
 The local spin susceptibility $\chi_{\rm L}(\Omega)$ is obtained 
from $\chi_{\rm L}({\rm i}\Omega_{n})$ through the analytic continuation.

\begin{figure}[ht]
\begin{center}
\includegraphics[width=6.2cm]{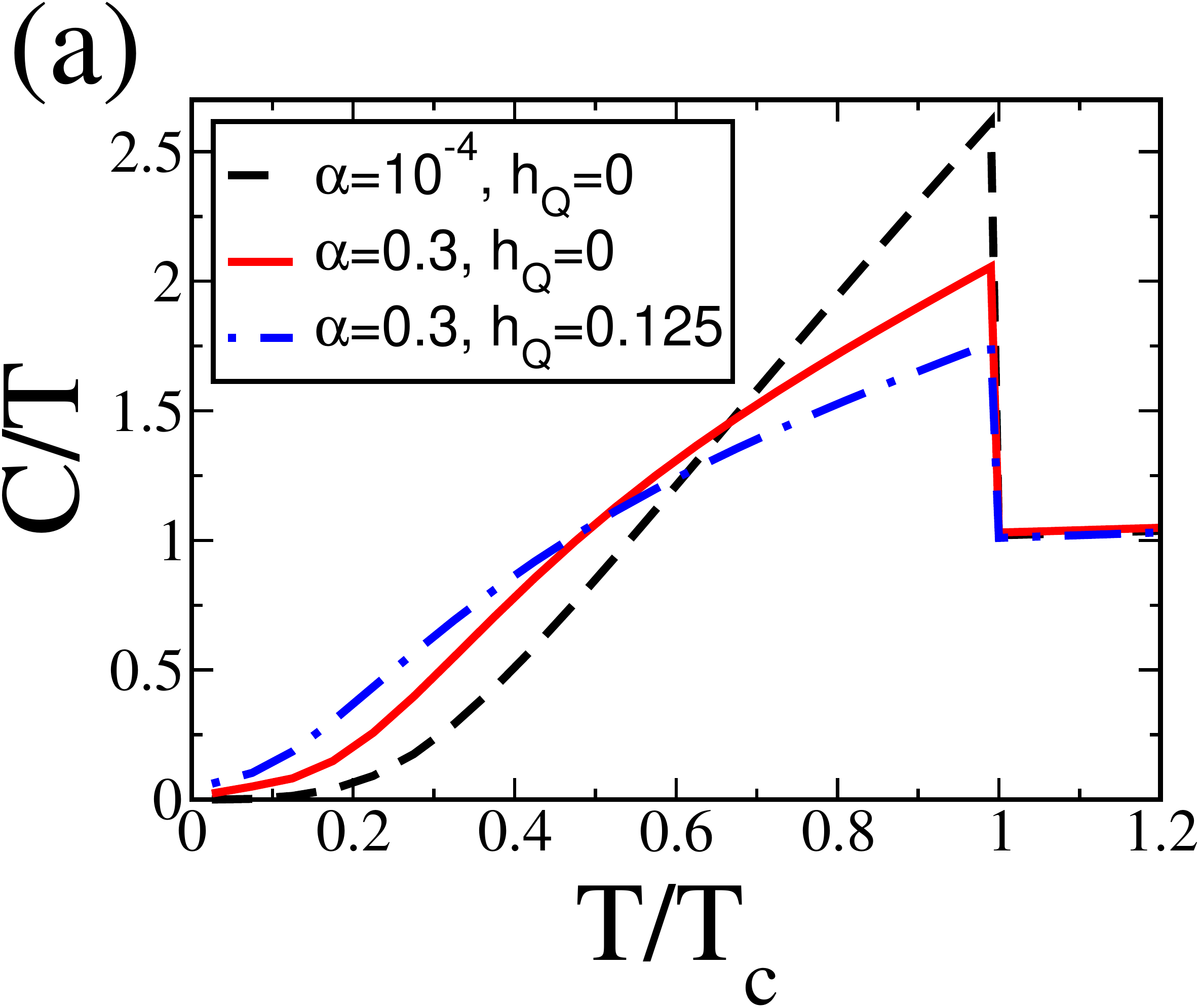}
\\
\vspace*{5mm}
\includegraphics[width=6.5cm]{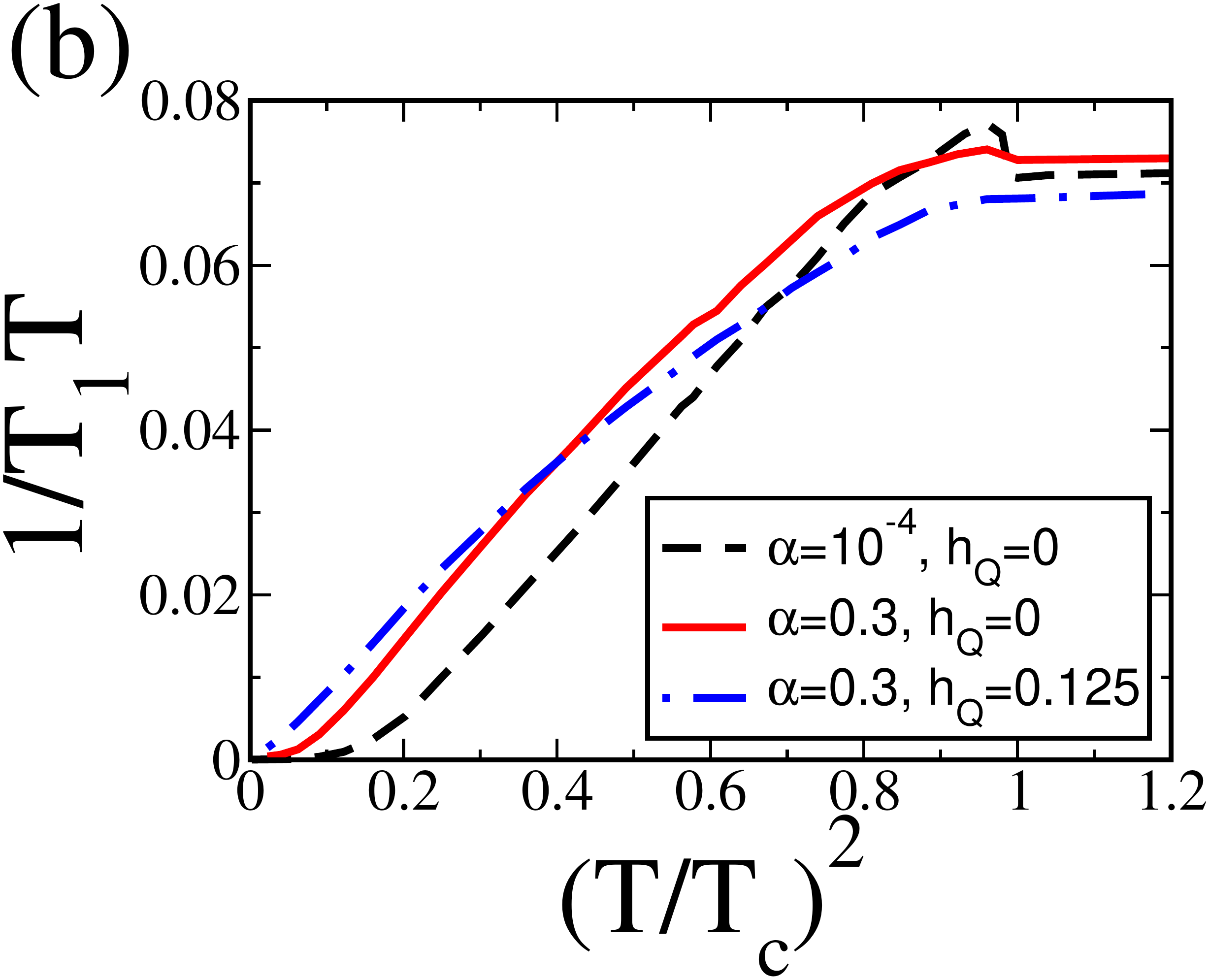}
\caption{(color online)
(a) Sommerfeld coefficient $C/T$ and 
(b) NMR $1/T_{1}T$ in the $s$+$P$-wave state. 
The parameters $\alpha$ and $h_{\rm Q}$ are shown in the figure. 
We solve the \eli equation in 64 $\times$ 64 $\times$ 32 lattices 
and estimate $C/T$ and $1/T_{1}T$ in 384 $\times$ 384 $\times$ 192 
lattices. 
} 
\end{center}
\end{figure}

 Figures~9(a) and 9(b) respectively show the temperature dependences of the 
Sommerfeld coefficient $C/T$ and the NMR $1/T_{1}T$ in the $s$+$P$-wave 
state. 
 When we assume a weak ASOC ($|\alpha| \ll T_{\rm c}$) and 
the absence of the AFM order (dashed lines in Fig.~9), 
both the Sommerfeld coefficient and the NMR $1/T_{1}T$ at low temperatures 
are much smaller than those expected in the superconductor with line nodes. 
 For example, the Sommerfeld coefficient shows a $T^{n}$ dependence 
($n > 2$) which is incompatible with the experimental 
result.~\cite{rf:takeuchiC} 
 On the other hand, we clearly see the line node behavior 
in the presence of the ASOC and AFM order (dash-dotted lines in Fig.~9). 
 The Sommerfeld coefficient obeys the 
$T$-linear law and the NMR $1/T_{1}T$ shows a $T^{2}$ dependence 
at low temperatures. These results are consistent with the 
experimental data of  specific heat,~\cite{rf:takeuchiC} 
thermal conductivity,~\cite{rf:izawa} superfluid density,~\cite{rf:bonalde} 
and NMR $1/T_{1}T$.~\cite{rf:mukudaT1}

 Upon decreasing the staggered field $h_{\rm Q}$, 
low-energy excitations are suppressed. 
 In the paramagnetic state ($h_{\rm Q}=0$), the Sommerfeld coefficient 
deviates from the $T$-linear law below $T < 0.2 T_{\rm c}$,  
while the $T^{2}$ dependence of NMR $1/T_{1}T$ breaks down at lower 
temperatures, $T < 0.1 T_{\rm c}$ (solid lines in Fig.~9). 
 If the AFM order is the main source of the line node in \Pt at 
ambient pressure, these deviations from the line node behavior may 
be observed at high pressures $P > 0.6$GPa.

\begin{figure}[ht]
\begin{center}
\includegraphics[width=6.2cm]{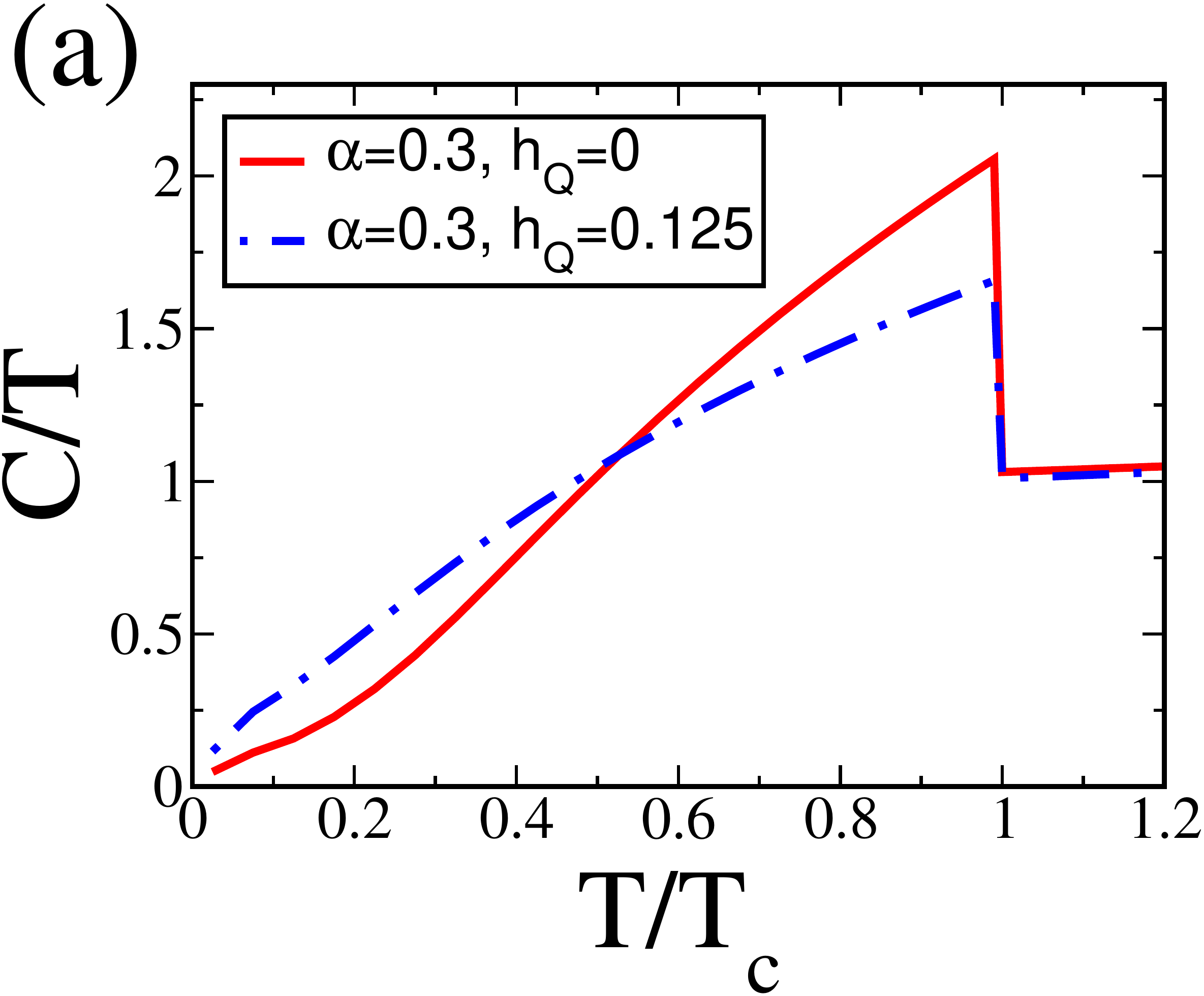}
\\
\vspace*{5mm}
\includegraphics[width=6.5cm]{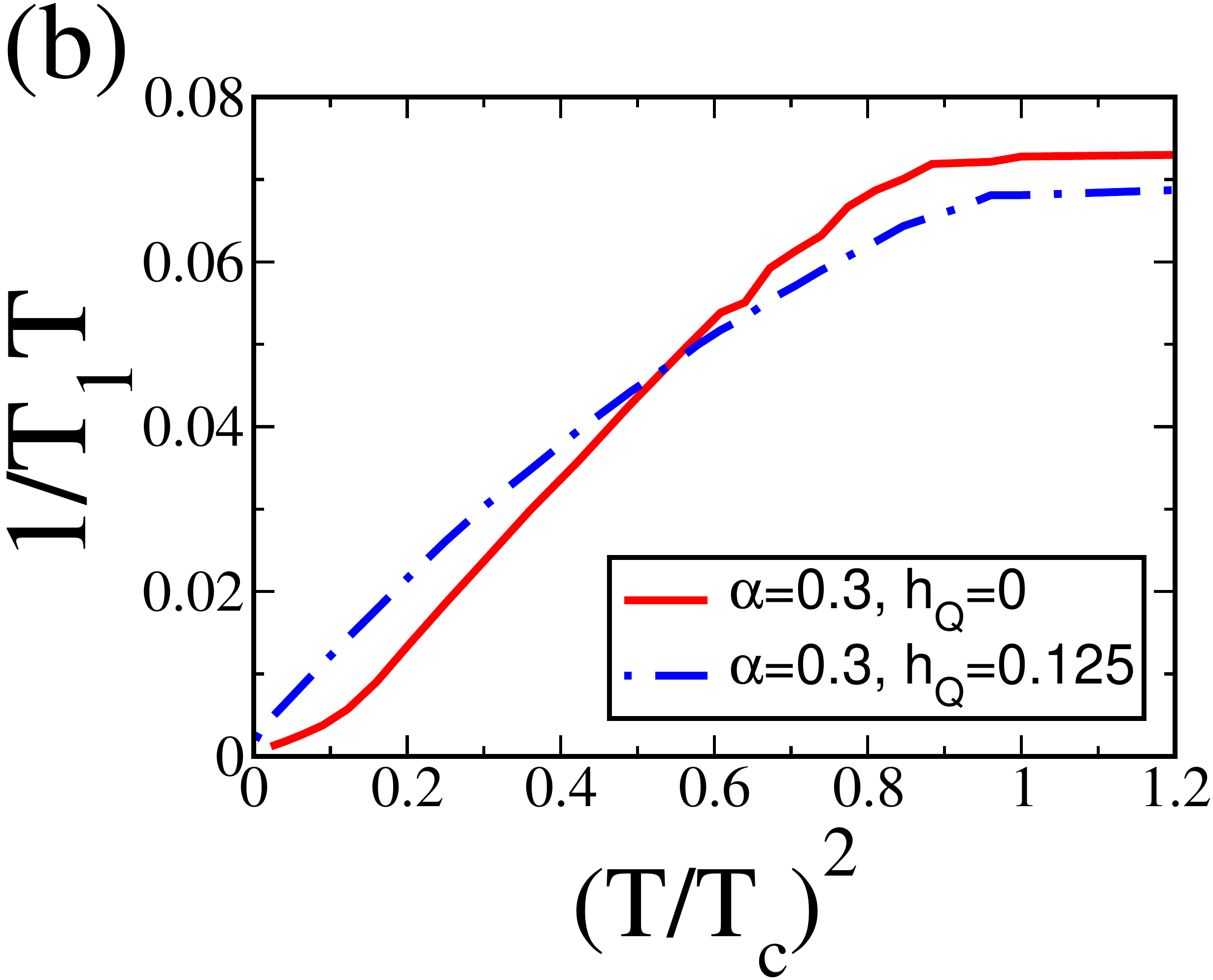}
\caption{(color online)
(a) Sommerfeld coefficient $C/T$ and 
(b) NMR $1/T_{1}T$ in the $p$+$D$+$f$-wave state. 
} 
\end{center}
\end{figure}

 Figure~10 shows the Sommerfeld coefficient and NMR $1/T_{1}T$ 
in the $p$+$D$+$f$-wave state. 
 The line node behavior appears clearly in both the paramagnetic and AFM states. 
 The role of the AFM order is qualitatively the same as that in the $s$+$P$-wave state: 
the number of low-energy excitations is increased by the AFM order. 
 This is because the vertical line node in the $d_{\rm xz}$-wave state 
disappears in the chiral $d$-wave state.

 We here discuss the coherence peak in the NMR $1/T_{1}T$. 
 It has been shown that the coherence peak appears in the $s$+$P$-wave state 
just below \Tc owing to the finite coherence 
factor.~\cite{rf:hayashiT1,rf:fujimoto} 
 This is the case in our calculation; however, the coherence peak is 
much smaller than that shown in ref.~30, as shown in Fig.~9(b).  
 This is because of the small ASOC $\alpha = 0.3 \ll \e_{\rm F}$ assumed in 
this paper and the extended $s$-wave nature of the spin singlet order parameter. 
 The coherence factor in the extended $s$-wave state is decreased by the 
sign reversal of the order parameter in the radial direction. 
Note that the isotropic $s$-wave pairing is generally not favored 
in the strongly 
correlated electron systems. 
 A slightly larger coherence peak appears in the paramagnetic state 
(solid line in Fig.~9(b)); however, this is not due to the coherence factor 
but arises from the anomaly in the DOS. 
 Although a coherence peak was reported in the early measurement of NMR 
$1/T_{1}T$,~\cite{rf:yogiT1} the recent measurement for a clean 
sample shows no coherence peak just below \Tcf,~\cite{rf:mukudaT1}
in agreement with our result.

\subsection{Multiple phase transitions}

 We have discussed the pressure dependence of low-energy excitations 
in \S4.2 and \S4.3. 
 Although qualitatively the same results are obtained for the 
low-energy excitations between the $s$+$P$-wave and $p$+$D$+$f$-wave states, 
there is an essential difference, 
namely, the multiple phase transitions in the $P$-$T$ plane. 
 To illustrate this issue, we show the possible phase diagrams in Fig.~11.

\begin{figure}[ht]
\begin{center}
\includegraphics[width=7cm]{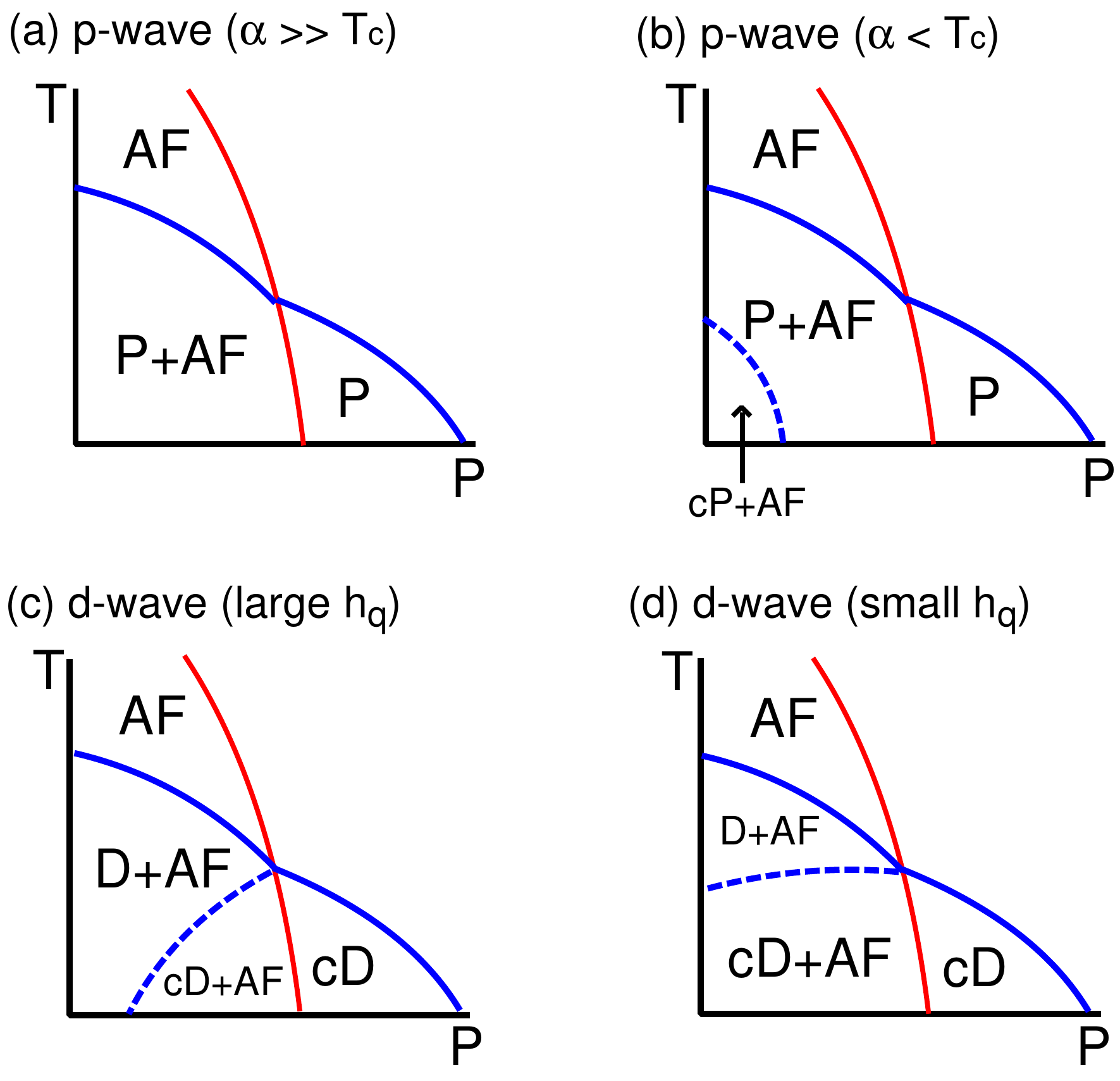}
\caption{(color online)
Possible phase diagrams in the $P$-$T$ plane. 
(a) $s$+$P$-wave state for a large ASOC ($|\alpha| \gg T_{\rm c}$). 
(b) $s$+$P$-wave state for a small ASOC ($|\alpha| \leq T_{\rm c}$). 
(c) $p$+$D$+$f$-wave state for a large staggered field $h_{\rm Q}$. 
(d) $p$+$D$+$f$-wave state for a small staggered field. 
``D'' (``cD'') shows the $d_{\rm xz}$-wave 
(chiral $d_{\rm xz} \pm {\rm i}d_{\rm yz}$-wave) state. 
``cP'' shows the chiral $p$-wave state where the dominant order parameter 
is $\vec{d}=(p_{\rm x} \pm {\rm i} p_{\rm y})\hat{x}$. 
} 
\end{center}
\end{figure}

 Figures~11(a) and 11(b) show the phase diagrams in the $s$+$P$-wave state. 
 When the ASOC is small ($|\alpha| \leq T_{\rm c}$), the chiral $p$-wave state 
is stabilized at low temperatures and low pressures, as in Fig.~11(b). 
 However, this is unlikely for CePt$_3$Si since the ASOC is much larger 
than \Tc in heavy fermion systems. 
 Therefore, the simple phase diagram in Fig.~11(a) is expected in the 
$s$+$P$-wave state of \Ptf.

 In the case of the $p$+$D$+$f$-wave state, the phase transition from the chiral 
$d_{\rm xz} \pm {\rm i} d_{\rm yz}$-wave state to the $d_{\rm xz}$-wave 
state must occur, as in Fig.~11(c) or 11(d). 
 When the staggered field  $h_{\rm Q}$ is large (small) at ambient pressure, 
the phase diagram in Fig.~11(c) (Fig.~11(d)) is expected. 
 Thus, the enhancement of the low-energy DOS due to pressure 
accompanies the second order phase transition, 
in contrast to the $s$+$P$-wave state.  
The observation of a multiple phase transition in the $P$-$T$ plane 
might provide clear evidence of the $p$+$D$+$f$-wave case.
 Although the second SC transition has been observed 
in \Ptf,~\cite{rf:nakatsuji,rf:bonalde} 
it has been shown that there are two SC phases with $T_{\rm c} \sim 0.75$K 
and $T_{\rm c} \sim 0.45$K in the sample.~\cite{rf:nakatsuji,rf:takeuchiC,
rf:aoki,rf:motoyama} 
 The second transition below \Tc seems to be caused by sample inhomogeneity.

\subsection{Anisotropy of upper critical field $H_{\rm c2}$}

 We here comment on the in-plane anisotropy of $H_{\rm c2}$ arising 
from the AFM order. 
 As discussed in \S4.2, the $p$-wave order parameter in the $s$+$P$-wave 
$\vec{d}(\k) \sim (-\sin k_{\rm y}, \beta \sin k_{\rm x},0)$ has a 
two-fold in-plane anisotropy in the AFM state. 
 The anisotropy parameter $\beta$ can be measured by the 
in-plane anisotropy of \Hc near \Tcf, which is determined by the orbital 
depairing effect and written as 
$H_{\rm c2}^{\rm a} = \Phi_{0}/(2 \pi \xi_{\rm b} \xi_{\rm c})$ and 
$H_{\rm c2}^{\rm b} = \Phi_{0}/(2 \pi \xi_{\rm a} \xi_{\rm c})$ for 
$\vec{H} \parallel \hat{a}$ and  $\vec{H} \parallel \hat{b}$, respectively.  
 Here, $\Phi_{0}=\frac{hc}{2e}$ is the flux quantum and 
$\xi_{\rm a,b,c}=\xi_{\rm a,b,c}^{0}(1-T/T_{\rm c})^{-1/2}$ 
are coherence lengths. 
 We obtain the ratio of the gradient 
$H_{\rm c2}^{'\rm a,b,c} = 
-T_{\rm c} {\rm d}H_{\rm c2}^{\rm a,b,c}/{\rm d}T$ as, 
$H_{\rm c2}^{'\rm a} : H_{\rm c2}^{'\rm b} : H_{\rm c2}^{'\rm c} 
= \xi_{\rm a}^{0} : \xi_{\rm b}^{0} : \xi_{\rm c}^{0}$ which 
can be estimated using the relation, 
$(\xi_{\rm a,b,c}^{0})^{2} \propto \sum'_{\gamma,k} 
v_{\gamma,\rm a,b,c}^{2}(\k) |\Psi_{\gamma}(\k)|^{2} 
f''(e_{\gamma}(\k))/8 e_{\gamma}(\k)$, where 
$v_{\gamma,\rm a,b,c}(\k) = {\rm d}e_{\gamma}(\k)/{\rm d}k_{\rm a,b,c}$ 
is the quasiparticle velocity in the $\gamma$-th band.

 Figure~12 shows the in-plane anisotropy 
$H_{\rm c2}^{'\rm a}/H_{\rm c2}^{'\rm b}$ in the $s$+$P$-wave state 
(solid line). 
 It is clearly shown that the anisotropy is induced by the AFM order 
for $h_{\rm Q} > 0.1$. This is mainly due to the decrease in the anisotropy 
parameter $\beta$. 
 Since $\beta < 1$ in the RPA theory and we assume the AFM staggered moment pointing along 
the $a$-axis,  \Hc is higher along the $b$-axis than along the $a$-axis 
($H_{\rm c2}^{\rm a} < H_{\rm c2}^{\rm b}$). 
 If $\beta > 1$, the opposite anisotropy appears. 
 Thus, if the marked mixing of $p$-wave order parameters due to the AFM order 
occurs, a pronounced in-plane anisotropy appears in \Hcf.

 The paramagnetic depairing effect qualitatively induces 
the same in-plane anisotropy as that in Fig.~12. 
 We have shown in ref.~29 a schematic figure 
of the $H$-$T$ phase diagram by taking into account both the orbital 
and paramagnetic depairing effects.

\begin{figure}[ht]
\begin{center}
\includegraphics[width=6cm]{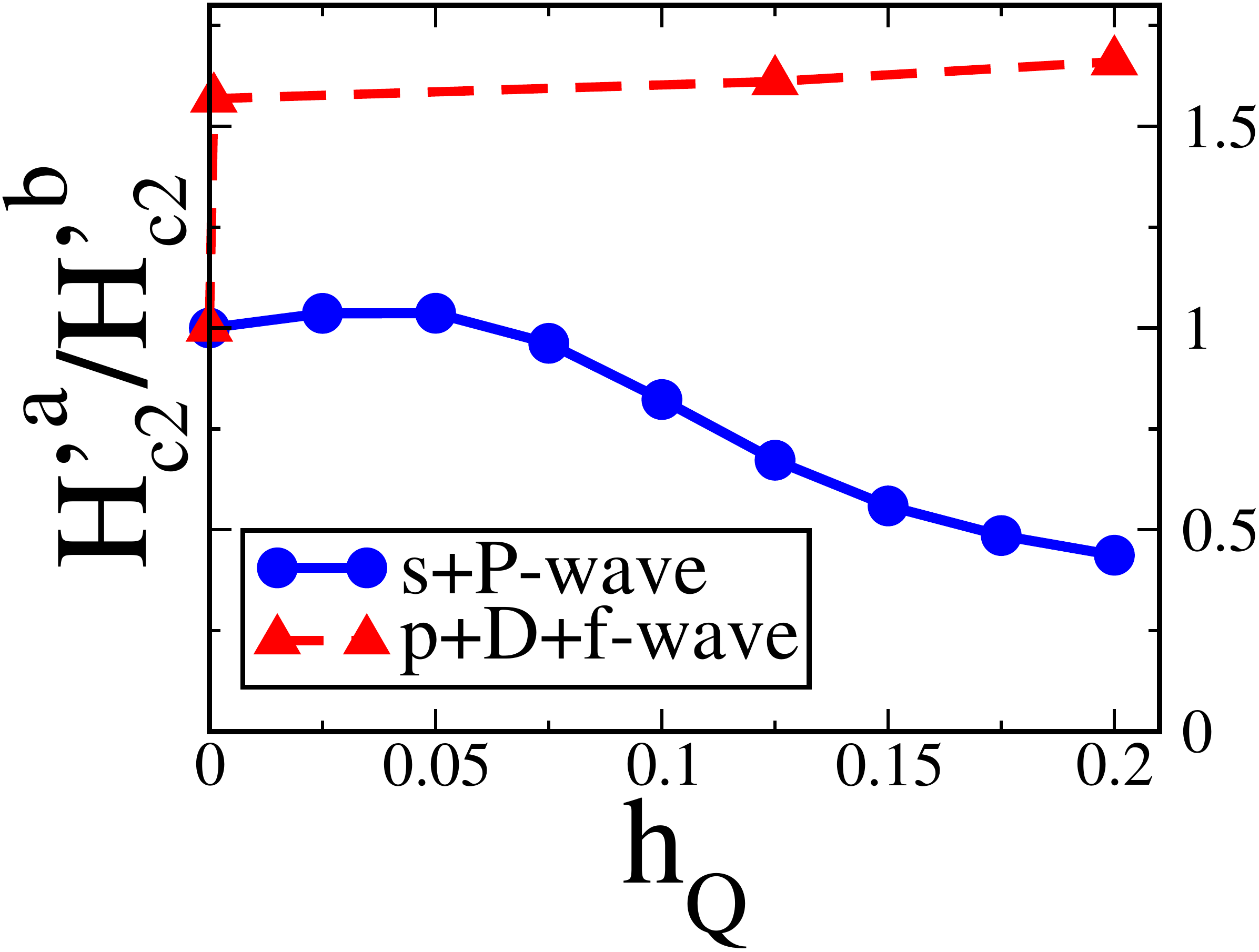}
\caption{(color online)
In-plane anisotropy of \Hc in the $s$+$P$-wave (solid line) 
and $p$+$D$+$f$-wave (dashed line) states. 
We assume $\alpha=0.3$. 
We show the ratio of the gradient $H_{\rm c2}^{'\rm a} = 
-T_{\rm c} {\rm d}H_{\rm c2}^{\rm a}/{\rm d}T$ and 
$H_{\rm c2}^{'\rm b} = -T_{\rm c} {\rm d}H_{\rm c2}^{\rm b}/{\rm d}T$ where 
$H_{\rm c2}^{\rm a}$ and $H_{\rm c2}^{\rm b}$ are the upper critical fields 
along {\it a}- and {\it b}-axes, respectively. 
} 
\end{center}
\end{figure}

 The in-plane anisotropy of \Hc in the $p$+$D$+$f$-wave state 
is quite different from that in the $s$+$P$-wave state. 
 We obtain $H_{\rm c2}^{'\rm a}/H_{\rm c2}^{'\rm b} \sim 1.6$ in the 
$p$+$D$+$f$-wave state independent of the nonzero staggered field $h_{\rm Q}$. 
 This two-fold anisotropy changes discontinuously way from 
$H_{\rm c2}^{'\rm a}/H_{\rm c2}^{'\rm b} \sim 1.6$ in the AFM state to 
$H_{\rm c2}^{'\rm a}/H_{\rm c2}^{'\rm b} = 1$ in the paramagnetic state, 
in contrast to the continuous change in the $s$+$P$-wave state. 
 It is therefore expected that the $s$+$P$-wave state can be distinguished 
from the $p$+$D$+$f$-wave state by the pressure dependence of 
in-plane anisotropy in \Hcf.

 Next we comment on the experimental measurement of in-plane anisotropy 
arising from the AFM order. 
 The direction of the AFM moment can be controlled by the cooling process, 
namely zero-field cooling and field cooling. 
 When temperature is decreased under the magnetic field along the 
$b$-axis, the AFM moment parallel to the $a$-axis appears 
below the \Neel temperature,  
because the system gains the maximum magnetic energy when 
the AFM moment is perpendicular to the magnetic field. 
 Then, the two-fold anisotropy due to the AFM order appears 
at low magnetic fields, although the AFM moment may rotate 
at high magnetic fields. 
 On the other hand, the domain structure with respect to the direction of 
AFM moment can appear when the system is cooled under a zero magnetic field. 
 Then the two-fold anisotropy is obscured.

 Before closing this section, some comments are given on the anisotropy of 
\Hc between the {\it ab}-plane and the {\it c}-axis. 
 We cannot discuss this anisotropy in a final way 
because not only the $\beta$-band but also the other band 
affects the anisotropy. 
 However, it should be noted that  \Hc is of similar magnitude along
the {\it ab}-plane and the {\it c}-axis because the $\beta$-band has a 
three-dimensional Fermi surface. 
 For example, we obtain 
$H_{\rm c2}^{'\rm a} : H_{\rm c2}^{'\rm b} : H_{\rm c2}^{'\rm c} 
= 1 : 1 : 0.71$ for $h_{\rm Q}=0$ and 
$H_{\rm c2}^{'\rm a} : H_{\rm c2}^{'\rm b} : H_{\rm c2}^{'\rm c} 
= 0.67 : 1 : 0.55$ for $h_{\rm Q}=0.125$ in the $s$+$P$-wave state, 
while 
$H_{\rm c2}^{'\rm a} : H_{\rm c2}^{'\rm b} : H_{\rm c2}^{'\rm c} 
= 1 : 1 : 0.85$ for $h_{\rm Q}=0$ and 
$H_{\rm c2}^{'\rm a} : H_{\rm c2}^{'\rm b} : H_{\rm c2}^{'\rm c} 
= 1.61 : 1 : 0.86$ for $h_{\rm Q}=0.125$ in the $p$+$D$+$f$-wave state. 
 The weak anisotropy of \Hc between $H \parallel ab$ and $H \parallel c$ 
is consistent with the experimental 
result for \Ptf.~\cite{rf:yasuda}

\section{Summary and Discussion}

 We have investigated the superconductivity in the Hubbard model 
with Rashba-type spin-orbit coupling and AFM order. 
 Applying the RPA theory to the $\beta$-band of CePt$_3$Si, we found two 
stable pairing states, the intraplane $p$-wave state 
admixed with the $s$-wave component ($s$+$P$-wave state) 
and the interplane $d$-wave state 
admixed with the $p$- and $f$-wave components ($p$+$D$+$f$-wave state). 
 We found that the anisotropy of helical spin fluctuation 
favors the $s$+$P$-wave state.

 We examined the low-energy excitations in detail. 
 The SC gap in the $p$+$D$+$f$-wave state has a line node protected by the 
symmetry, while accidental line nodes appear in the $s$+$P$-wave state. 
 Thus, both pairing states seem to be consistent with the experimental 
results in \Pt at ambient 
pressure.~\cite{rf:takeuchiC, rf:bonalde,rf:izawa,rf:mukudaT1} 
 A substantial part of the accidental line node in the $s$+$P$-wave state 
can be induced by the AFM order through the rotation of the $d$-vector. 
 The line node in the $p$+$D$+$f$-wave state is also increased by the 
AFM order because of the phase transition from the chiral $d$-wave state 
in the paramagnetic state to the $d_{\rm xz}$-wave state in the AFM state. 
 Thus, the number of low-energy excitations decreases in both states 
when the AFM order is suppressed by pressure. 
 We calculated the specific heat and NMR $1/T_{1}T$  in both the 
paramagnetic and AFM states. 
 The deviation from the line node behavior in the paramagnetic state 
has been pointed out.

 We proposed some future experiments that can elucidate the pairing 
state in \Ptf. 
The first one is the pressure dependence of low-energy excitations 
discussed above. 
 Another one is the possible multiple SC phase transitions 
in the $P$-$T$-plane. 
 The second SC transition occurs below \Tc near the critical pressure 
for the AFM order, if the $p$+$D$+$f$-wave superconductivity is realized. 
 This is in contrast to the $s$+$P$-wave state where 
no additional phase transition is expected. 
 The marked change of low-energy excitations in the $p$+$D$+$f$-wave state 
is accompanied by the second order phase transition. 
 The last proposal is the anisotropy of \Hc in the {\it ab}-plane. 
 In the $s$+$P$-wave state, the anisotropy of \Hc gradually increases 
with increasing AFM moment, while that in the $p$+$D$+$f$-wave state 
is discontinuous at a critical pressure for the AFM order. 
 Our proposals for future experiments do not rely on the particular 
band structure of the $\beta$-band in \Ptf, and therefore can also be applied to 
CeRhSi$_3$, CeIrSi$_3$, and CeCoGe$_3$.

 According to the present experiments, 
the $s$+$P$-wave superconductivity is most likely realized in \Ptf. 
 The paramagnetic properties measured on the basis of the NMR Knight shift and 
\Hc seem to be compatible with those in the $s$+$P$-wave state.~\cite{rf:yanasehelical}  
 Futher studies from both the theoretical and experimental points of view 
are highly desired to elucidate the novel physics 
in the non-centrosymmetric superconductivity.

\section*{Acknowledgments}

 The authors are grateful to D. F. Agterberg, J. Akimitsu, S. Fujimoto, 
J. Flouquet, N. Hayashi, R. Ikeda, K. Izawa, N. Kimura, Y. Kitaoka, 
Y. Matsuda, M.-A. Measson, V. P. Mineev, G. Motoyama, H. Mukuda, Y. Onuki, 
T. Shibauchi, R. Settai, T. Tateiwa, and M. Yogi for fruitful discussions. 
 This study was financially supported by the Nishina Memorial 
Foundation, Grants-in-Aid for Young Scientists (B) from MEXT, Japan, 
Grants-in-Aid for Scientific Research on Priority Areas (No. 17071002) 
from MEXT, Japan, the Swiss Nationalfonds, and the NCCR MaNEP. 
 Numerical computation in this work was carried out 
at the Yukawa Institute Computer Facility.

\appendix

\section{Derivation of Rashba-type ASOC in Tight-binding Models}

 We here microscopically derive the Rashba-type spin-orbit coupling 
in the periodical Anderson model and Hubbard model by 
tight-binding approximation. 

 The localized 4$f$ states in the Ce-based heavy fermion superconductors, 
such as \Ptf, \Rhf, and \Ir are described by the $J=5/2$ manifold 
whose degeneracy is split by the crystal electric field.  
 The 4$f$ levels in \Pt are described by the 
three doublets,~\cite{rf:commentCEF} 
\begin{eqnarray}
&& \hspace*{-10mm}
|\Gamma_{7} \pm>  =
\sqrt{\frac{5}{6}} |\pm \frac{5}{2}> 
- \sqrt{\frac{1}{6}} |\mp \frac{3}{2}>,
\\ && \hspace*{-10mm}
|\Gamma'_{6} \pm>  = |\pm \frac{1}{2}>, 
\\ && \hspace*{-10mm}
|\Gamma'_{7} \pm> =
\sqrt{\frac{1}{6}} |\pm \frac{5}{2}> 
+ \sqrt{\frac{5}{6}} |\mp \frac{3}{2}>. 
\end{eqnarray}
 The ground state is $|\Gamma_{7} \pm>$, and the excited 
$|\Gamma'_{6} \pm>$ and $|\Gamma'_{7} \pm>$ states 
have excitation energies of  $1$ and $24$meV, respectively.~\cite{rf:metoki}

 Next we construct a periodical Anderson model for the 
$|\Gamma_{7} \pm>$ state, which hybridizes with conduction 
electrons. 
 It is straightforward to apply the following procedure to the  
$|\Gamma'_{6} \pm>$ and $|\Gamma'_{7} \pm>$ states. 
 Because the mirror symmetry is broken along the {\it z}-axis in 
\Ptf, the odd parity Ce 4$f$-orbital is hybridized with the even parity 
$s$- and $d$-orbitals in the same Ce site. 
 Owing to the symmetry of the  $|\Gamma_{7} \pm>$ state, 
the localized 4$f$ state is hybridized with the $d_{\rm xy}$-, 
$d_{\rm xz}$- and $d_{\rm yz}$-orbitals. 
 Then, the wave function of the localized state can be expressed as, 
\begin{eqnarray}
&& \hspace*{-20mm}
|{\rm f} \pm> =
\kappa |\Gamma_{7} \pm> + {\rm i} 
\epsilon |{\rm d}_{\rm xy}> \chi_{\pm} 
\nonumber \\ && \hspace*{-10mm}
+ \eta (|{\rm d}_{\rm xz}> 
\mp {\rm i} |{\rm d}_{\rm yz}>) \chi_{\mp}, 
\end{eqnarray}
where $\epsilon$, $\eta$ and $\kappa=\sqrt{1-\epsilon^{2}-2 \eta^{2}}$ 
are real and $\chi_{\pm}$ describes the wave function of the spin. 
 We note that the wave function of $|\Gamma_{7} \pm>$ 
is given by 
\begin{eqnarray}
&& \hspace*{-15mm}
|\Gamma_{7} \pm> = 
- \sqrt\frac{5}{21} {\rm i} |L_{\rm z}=2-> \chi_{\pm}
\nonumber \\ && \hspace*{-10mm}
\pm (\sqrt\frac{15}{21} |L_{\rm z}=\pm 3> 
- \sqrt\frac{1}{21} |L_{\rm z}= \mp 1>) \chi_{\mp}, 
\end{eqnarray}
where $|L_{\rm z}=2-> = \frac{1}{\sqrt{2}{\rm i}} 
(|L_{\rm z}=2> - |L_{\rm z}=-2>)$.

 The periodical Anderson Hamiltonian is constructed for the localized 
$|{\rm f} \pm>$ state and  conduction electrons. 
 We here consider the conduction electrons arising from the 
Ce 5$s$-orbital for simplicity. 
 Taking into account the inter-site hybridization between the 
$s$-, $d$- and $f$-orbitals, we obtain the tight-binding 
Hamiltonian 
\begin{eqnarray}
&& \hspace*{-40mm}
H_{0} = \sum_{\k} \hat{\psi}_{\k}^{\dag} \hat{H}_{0}(\k) \hat{\psi}_{\k}, 
\end{eqnarray}
where $\hat{\psi}_{\k}^{\dag} = (f^{\dag}_{\k +}, f^{\dag}_{\k -}, 
c^{\dag}_{\k \uparrow}, c^{\dag}_{\k \downarrow})$, and 
\begin{eqnarray}
&& \hspace*{-30mm}
\hat{H}_{0}(\k)
= \left(
\begin{array}{cc}
\hat{\e}_{\rm f}(\k) & \hat{V}(\k) \\
\hat{V}(\k)^{\dag} & \hat{\e}_{\rm c}(\k) \\
\end{array}
\right). 
\end{eqnarray}
 The 2 $\times$ 2 matrix $\hat{\e}_{\rm f}(\k)$, $\hat{\e}_{\rm c}(\k)$, 
and $\hat{V}(\k)$ are obtained as 
\begin{eqnarray}
&&
\hat{\e}_{\rm f}(\k)
= \left(
\begin{array}{cc}
\e_{\rm f}(\k) & \alpha_{1} ({\rm i} {\rm s}_{\rm x} +{\rm s}_{\rm y})  \\
\alpha_{1} (-{\rm i} {\rm s}_{\rm x} +{\rm s}_{\rm y}) & \e_{\rm f}(\k) \\
\end{array}
\right), 
\\ && 
\hat{\e}_{\rm c}(\k)
= \left(
\begin{array}{cc}
\e_{\rm c}(\k) & 0 \\
0& \e_{\rm c}(\k) \\
\end{array}
\right), 
\\ &&
\hat{V}(\k) = 
\nonumber \\ && \hspace*{-10mm}
\left(
\begin{array}{cc}
(8 V_{3} {\rm s}_{\rm z} + 4 {\rm i} \epsilon V_{4})
{\rm s}_{\rm x} {\rm s}_{\rm y} 
& (2 {\rm i} V_{5} - 4 \eta V_{6} {\rm s}_{\rm z})
({\rm s}_{\rm x} + {\rm i} {\rm s}_{\rm y}) \\
(2 {\rm i} V_{5} - 4 \eta V_{6} {\rm s}_{\rm z})
({\rm s}_{\rm x} - {\rm i} {\rm s}_{\rm y}) 
& (8 V_{3} {\rm s}_{\rm z} + 4 {\rm i} \epsilon V_{4}) 
{\rm s}_{\rm x} {\rm s}_{\rm y} \\
\end{array}
\right), 
\nonumber \\ 
\end{eqnarray}
where the abbreviation ${\rm s}_{\rm x,y,z} = \sin k_{\rm x,y,z}$ is used. 
 We ignored the off-diagonal terms in the second order 
with respect to the small parameters $\epsilon$ and $\eta$. 
 We obtain $\e_{\rm f}(\k) 
= \kappa^{2} \e_{\Gamma_{7}}(\k) 
+ \epsilon^{2} \e_{\rm xy}(\k) + \eta^{2} (\e_{\rm xz}(\k) + \e_{\rm yz}(\k))$ 
where $\e_{\rm A}(\k)$ is the dispersion relation for 
the $|{\rm A}>$ state. 
 It is clearly shown that eq.~(A.8) has the Rashba type spin-orbit coupling 
term and the coefficient is obtained as 
\begin{eqnarray}
&& \hspace*{-20mm}
\alpha_{1} = -4 \epsilon V_{2} - 4 \eta V_{1}. 
\end{eqnarray} 
 The hybridization parameters in eqs.~(A.10) and (A.11) are obtained as 
\begin{eqnarray}
&& \hspace{-10mm}
V_{1} = \kappa \sqrt\frac{5}{21} V^{100}_{\rm 2-,yz}, 
\\ && \hspace{-10mm}
V_{2} = \kappa 
(\sqrt\frac{15}{42} V^{100}_{\rm y^{3}-3 x^{2}y,yz} 
-\sqrt\frac{1}{42} V^{100}_{\rm y(5z^{2}-r^{2}),yz}),
\\ && \hspace{-10mm}
V_{3} = \kappa \sqrt\frac{5}{21} V^{111}_{\rm 2-,s},
\\ && \hspace{-10mm}
V_{4} = V^{110}_{\rm xy,s}, 
\\ && \hspace{-10mm}
V_{5} = \kappa 
(\sqrt\frac{15}{42} V^{100}_{\rm x^{3}-3 xy^{2},s} 
-\sqrt\frac{1}{42} V^{100}_{\rm x(5z^{2}-r^{2}),s}),
\\ && \hspace{-10mm}
V_{6} = V^{101}_{\rm xz,s}, 
\end{eqnarray}
where $V^{\rm abc}_{\rm A,B}$ is the hopping matrix element between  
the $|{\rm A}>$ and $|{\rm B}>$ states along 
the [abc]-axis. 

 Note that the parameters $\epsilon$ and $\eta$ arise from 
the intra-site hybridization between the $d$- and $f$-orbitals 
while the matrix elements $V_{1}$ and $V_{2}$ describe the inter-site 
hybridization. 
 Thus, the intra-orbital Rashba-type spin orbit coupling $\alpha_{1}$ 
arises from the hybridization of the $\Gamma_{7}$-state with the $d_{\rm xy}$-, 
$d_{\rm xz}$-, and $d_{\rm yz}$-states. 
 Note again that the parameters $\epsilon$ and $\eta$ vanish 
in centrosymmetric systems.

 Applying an appropriate unitary transformation to the conduction electron, 
$(c^{\dag}_{\k +}, c^{\dag}_{\k -}) = 
(c^{\dag}_{\k \uparrow}, c^{\dag}_{\k \downarrow}) \hat{U}_{\rm c}(\k)$, 
the hybridization matrix is transformed as  
\begin{eqnarray}
&& \hspace*{-10mm}
\tilde{V}(\k) 
= \hat{V}(\k) \hat{U}_{\rm c}(\k) =
\nonumber \\ && \hspace*{-18mm}
 \left(
\begin{array}{cc}
V_{\rm cf}(\k) 
& \alpha_{2}(\k) ({\rm i} {\rm s}^{2}_{\rm y} {\rm s}_{\rm x} 
                  + {\rm s}^{2}_{\rm x} {\rm s}_{\rm y}) \\
\alpha_{2}(\k) (-{\rm i} {\rm s}^{2}_{\rm y} {\rm s}_{\rm x} 
                  + {\rm s}^{2}_{\rm x} {\rm s}_{\rm y})
& V_{\rm cf}(\k) \\
\end{array}
\right), 
\end{eqnarray}
where 
\begin{eqnarray}
&& \hspace*{-18mm}
\alpha_{2}(\k)= 
4(\epsilon V_{4} V_{5} - 4 \eta V_{3} V_{6} {\rm s}_{\rm z}^{2})
\nonumber \\ && \hspace*{-3mm}
/\sqrt{16 V_{3}^{2} {\rm s}_{\rm x}^{2} {\rm s}_{\rm y}^{2} 
{\rm s}_{\rm z}^{2} + V_{5}^{2} ({\rm s}_{\rm x}^{2} + {\rm s}_{\rm y}^{2})}.
\end{eqnarray}
 Note that $\alpha_{2}(\k)$ is a real and even function with respect to 
$k_{\rm x}$, $k_{\rm y}$, and $k_{\rm z}$. 

 Taking into account the on-site repulsion in the 
$|{\rm f} \pm>$ state, 
we obtain the periodical Anderson model with a Rashba-type ASOC as 
\begin{eqnarray}
&& \hspace*{-15mm}  
H = H_{\rm k} + H_{\rm ASOC} + H_{\rm I}, 
\\ && \hspace*{-15mm}  
H_{\rm k} = 
     \sum_{k,s=\pm} \e_{\rm f}(\k) f_{\k,s}^{\dag}f_{\k,s} 
   + \sum_{k,s=\pm} \e_{\rm c}(\k) c_{\k,s}^{\dag}c_{\k,s} 
\nonumber \\ && \hspace*{-5mm}  
   + \sum_{k,s=\pm} [V_{\rm cf}(\k) f_{\k,s}^{\dag}c_{\k,s} + h.c.], 
\\ && \hspace*{-15mm}  
H_{\rm ASOC} = 
     \alpha_{1} \sum_{k,s,s'} \vec{g}_{\rm f}(\k) \cdot \vec{\sigma}_{ss'} 
                            f_{\k,s}^{\dag}f_{\k,s'} 
\nonumber \\ && \hspace*{-5mm}  
+ \sum_{k,s,s'} [\alpha_{2}(\k) \vec{g}_{\rm cf}(\k) \cdot \vec{\sigma}_{ss'} 
                            f_{\k,s}^{\dag}c_{\k,s'} + h.c.], 
\\ && \hspace*{-15mm}  
H_{\rm I} = 
     U \sum_{i} n^{\rm f}_{i,+} n^{\rm f}_{i,-}, 
\end{eqnarray}
where $\vec{g}_{\rm f}(\k)=(\sin k_{\rm y}, -\sin k_{\rm x},0)$ and 
$\vec{g}_{\rm cf}(\k)=(\sin^{2} k_{\rm x} \sin k_{\rm y}, 
-\sin^{2} k_{\rm x} \sin k_{\rm x},0)$ describe the $g$-vector for the 
intra- and inter-orbital Rashba-type ASOCs, respectively. 

 Note again that the ASOC arises from the 
atomic $L$-$S$ coupling in the Ce 4$f$-orbital and the parity mixing 
in the localized state. 
 The breakdown of the inversion symmetry plays an essential role in  
the parity mixing in the atomic state. 
 The inter-site hybridization between the $f$- and 
admixed $d$-(or $s$-)orbitals gives rise to the intra-orbital 
ASOC, while the inter-orbital ASOC is induced by 
the hybridization between the conduction electrons and the 
admixed $d$- (or $s$-) orbitals. 
 Note that the cubic term 
$\propto \sin k_{\rm x} \sin k_{\rm y} \sin k_{\rm z} 
(\sin^{2} k_{\rm x}-\sin^{2} k_{\rm y}) \sigma_{\rm z}$~\cite{rf:samokhinChi} 
does not appear in the above derivation.

 We derived the periodical Anderson model for the localized 
$|\Gamma'_{6} \pm> = |\pm \frac{1}{2}>$ 
state and the conduction electrons with $s$-orbital symmetry. 
 Then, we obtained the Hamiltonian that is similar to eq.~(A.20), but the 
$g$-vector for the inter-orbital ASOC is replaced with  
$\vec{g}_{\rm cf}(\k)=(\sin k_{\rm y},- \sin k_{\rm x},0) 
= \vec{g}_{\rm f}(\k)$. 
  Other crystal field levels different from eqs.~(A.1)-(A.3) 
have been proposed.~\cite{rf:bauerreview,rf:motoyamaLT}
 The Rashba-type ASOC can also be derived  
for these levels in the same way as above. 
 Thus, the Rashba-type ASOC is generally derived in the periodical 
Anderson Hamiltonian by taking into account the parity mixing 
in the atomic $4f$-state.

 The kinetic energy term $H_{\rm k}$ in the periodical Anderson model 
is diagonalized by the unitary transformation 
$(a^{\dag}_{1,\k \pm}, a^{\dag}_{2,\k \pm}) = 
(f^{\dag}_{\k \pm}, c^{\dag}_{\k \pm}) \hat{U}_{\rm cf}(\k)$ with 
\begin{eqnarray}
&& \hspace*{-15mm}
\hat{U}_{\rm cf}(\k) 
= 
 \left(
\begin{array}{cc}
a_{1}(\k) & a^{*}_{2}(\k) \\
a_{2}(\k) & -a_{1}(\k) \\
\end{array}
\right). 
\end{eqnarray}
 Applying this unitary transformation to the periodical Anderson model 
in eq.~(A.20) and dropping the upper band described by 
$a^{\dag}_{2,\k \pm}$, we obtain the single-orbital model with 
the Rashba-type ASOC. The $g$-vector is obtained as 
\begin{eqnarray}
&& \hspace*{-10mm}
\alpha \vec{g}(\k)= \alpha_{1} a_{1}(\k)^{2} \vec{g}_{\rm f}(\k) + 
\nonumber \\ && \hspace*{1mm}
             \alpha_{2}(\k)  a_{1}(\k)(a_{2}(\k)+a_{2}^{*}(\k)) 
             \vec{g}_{\rm cf}(\k). 
\end{eqnarray}
 The unitary transformation described by $U_{\rm cf}(\k)$ 
leads to the momentum dependence of the two-body interaction term 
$H_{\rm I}$, as in the case of the multi-orbital Hubbard model.~\cite{rf:yanaseCo} 
 By neglecting this momentum dependence for simplicity, we obtain the 
single-orbital Hubbard model in eq.~(1) where the $g$-vector is described 
by eq.~(A.25). 
 The investigation of the periodical Anderson model in eq.~(A.20) is an 
interesting future issue.


\begin{thebibliography}{9}
\bibitem{rf:commentCEF}
 The coefficients in $|\Gamma_{7} \pm>$ and 
$|\Gamma'_{7} \pm>$ 
are slightly different from those reported in ref.~57, 
but the difference is not important in the following discussion.
\end{thebibliography}
\end{document}